\documentclass[journal=jacsat,manuscript=article]{achemso}

\usepackage{titlesec}
\usepackage{amsmath, amsfonts, amssymb}
\usepackage{mathtools}
\usepackage[version=3]{mhchem} 
\usepackage[left]{lineno}
\usepackage[hidelinks]{hyperref}
\usepackage{xcolor}

\author{Antonio Grimaldi}
\affiliation[unibo]{Department of Physics and Astronomy, University of Bologna}
\author{Michele Stofella}
\affiliation[leeds]{The Astbury Centre for Structural Molecular Biology, University of Leeds}
\author{Billy Hobbs}
\affiliation[icl]
{Department of Life Sciences, Imperial College London}
\author{Theodoros K. Karamanos}
\affiliation[icl]
{Department of Life Sciences, Imperial College London}
\author{Emanuele Paci}
\email{e.paci@unibo.it}
\affiliation[unibo]{Department of Physics and Astronomy, University of Bologna}

\title[Amide Hydrogen Deuterium Exchange in Isotopically Mixed Waters]
  {Amide Hydrogen Deuterium Exchange in Isotopically Mixed Waters}

\begin{document}


\setstretch{1}


\begin{abstract}
Hydrogen–deuterium exchange (HDX) of protein backbone amides provides a powerful probe of conformational dynamics. However, when experiments are performed in \ce{H2O}/\ce{D2O} mixtures, quantitative interpretation is hindered by back exchange and isotope effects not captured by the classical Linderstrøm–Lang (LL) model. We introduce a generalized Linderstrøm–Lang (GLL) framework that explicitly accounts for forward and reverse exchange and for changes in protection upon isotopic substitution. Analytical solutions describe equilibrium enrichment (fractionation) and protection factors in mixtures, reducing to the LL model in pure \ce{D2O}. Application to HDX/NMR of the molecular chaperone DNAJB1 in 50\% \ce{D2O} demonstrates that the GLL model recovers protection factors at 100\% \ce{D2O}. Ignoring back exchange (i.e., using the LL model) protection factors are systematically underestimated. 
A particularly powerful feature of our approach is that a single HDX experiment in a mixture (e.g., 50\% \ce{D2O} ) simultaneously provides protection factors that report on conformational dynamics and local stability and fractionation factors that are sensitive to the local hydrogen-bonding environment.

\end{abstract}
\section*{Introduction}
\label{sec:intro}

Hydrogen deuterium exchange (HDX) is a spontaneous process in which the hydrogen atoms of a solute molecule are replaced with deuterium from solvent.
HDX in proteins was pioneered by Linderstr{\o}m-Lang and coworkers \cite{hvidt1954exchange, linderstrom1955ph, englander1997hydrogen} with the goal of measuring the exchange rates of backbone amides, which depend on protein primary and higher-order structure, conformational dynamics as well as physico-chemical properties of the solvent \cite{hvidt1966hydrogen, englander1983hydrogen, hamuro2020tutorial, james2021advances}.
Intrinsic HDX rates, that pertain to maximally solvated amides, have been measured in pure \ce{H2O} and \ce{D2O} \cite{bai1993primary, connelly1993isotope, nguyen2018reference}.
Observed HDX rates can be orders of magnitude lower than intrinsic ones, e.g.\! for amides engaged in stabilizing intramolecular hydrogen bonds or buried in a hydrophobic core. 
Effects of this kind, collectively termed protection \cite{hvidt1966hydrogen, englander1983hydrogen}, make HDX-based techniques suitable for fingerprinting protein structure and dynamics, with applications spanning the study of folding\cite{englander2016protein} and allostery\cite{englander2023hx} to protein-ligand interactions, and the development of novel therapeutics \cite{masson2017overview, krishnamurthy2025evolving}.
HDX data can aid prediction of protein conformational ensembles\cite{devaurs2022computational} and understanding of intrinsically disordered proteins \cite{ghafouri2025unifiedframeworkdeterminingconformational}.

The Linderstr{\o}m-Lang (LL) model \cite{hvidt1966hydrogen} describes HDX of proteins in pure \ce{D2O}.
It assumes that each backbone amide hydrogen adopts either closed (\ce{H_{cl}}) or open (\ce{H_{op}}) conformations, only the latter being competent to exchange according to the reaction
\begin{align}
    \ce{H_{cl}} \xrightleftharpoons[k_\mathrm{cl}]{k_\mathrm{op}} \ce{H_{op}} \xrightarrow{k_\mathrm{int}} \ce{exchanged}.
\label{eq:ll}
\end{align}
The rate constants of opening and closing transitions, $k_\mathrm{op}$ and $k_\mathrm{cl}$, encode protein conformational dynamics.
Their ratio is the protection factor $P = k_\mathrm{cl}/k_\mathrm{op}$, which is the reciprocal of the opening equilibrium constant, related to the opening free energy $\Delta G_\mathrm{op} = G_\mathrm{op} - G_\mathrm{cl}$ by\cite{hamuro2020tutorial}
\begin{align}
    \Delta G_\mathrm{op} = RT \ln P.
\label{eq:ll-DeltaG=RTlnP}
\end{align}
The intrinsic exchange rate $k_\mathrm{int}$ of an amide depends on  neighboring residues, temperature and pH \cite{bai1993primary,nguyen2018reference}.
At near-neutral pH, the conditions $k_\mathrm{cl} \gg k_\mathrm{op}$ ($P \gg 1$) and $k_\mathrm{cl} \gg k_\mathrm{int}$ are satisfied by most amides of native proteins, and exchange occurs in the so-called EX2 limit \cite{hvidt1966hydrogen}.
In this case, the exchanged fraction of initially undeuterated amides is given by a single exponential 
\begin{align}
    D(t) = 1 - \mathrm{e}^{-k_\mathrm{obs}t},
\label{eq:ll-single-exp}
\end{align}
with observed rate constant
\begin{align}
    k_\mathrm{obs} = \frac{k_\mathrm{int}}{P}.
\label{eq:ll-ex2-kobs}
\end{align}


A number of analytical methods sensitive to the properties of hydrogen isotopes can detect exchange \cite{engen2015analytical}.
The experiment (or steps thereof) is often performed in \ce{H2O}/\ce{D2O} mixtures.
The LL model \eqref{eq:ll} does not account for the back exchange and isotope effects observed in mixtures.
A generalized theoretical framework is developed here that incorporates these effects as an extension to the LL model, also opening the door to understanding (and, crucially, correcting for) back-exchange in HDX/MS workflows \cite{stofella2024computational, konermann-opportunities}.
The generalized model is applied to HDX/NMR measurements of \ce{^{15}N}-DNAJB1 JD-GF-${\rm \alpha 5}$ F94L performed in 50\% \ce{D2O}.
Protection factors extracted from the experiment in the mixture using the generalized model are consistent with the results for pure \ce{D2O} analyzed using the LL model.
An extra piece of information, solely available from measurements in mixtures, is the fractionation that reports on the local hydrogen bonding network\cite{cleland1994low, loh1994hydrogen, edison1995theoretical, bowers1996hydrogen, liwang1996equilibrium, krantz2000d} and may complement the protection factors in modeling structural ensembles.

\section*{Theoretical framework}

\begin{figure}
    \centering
    \includegraphics[width=.9\linewidth]{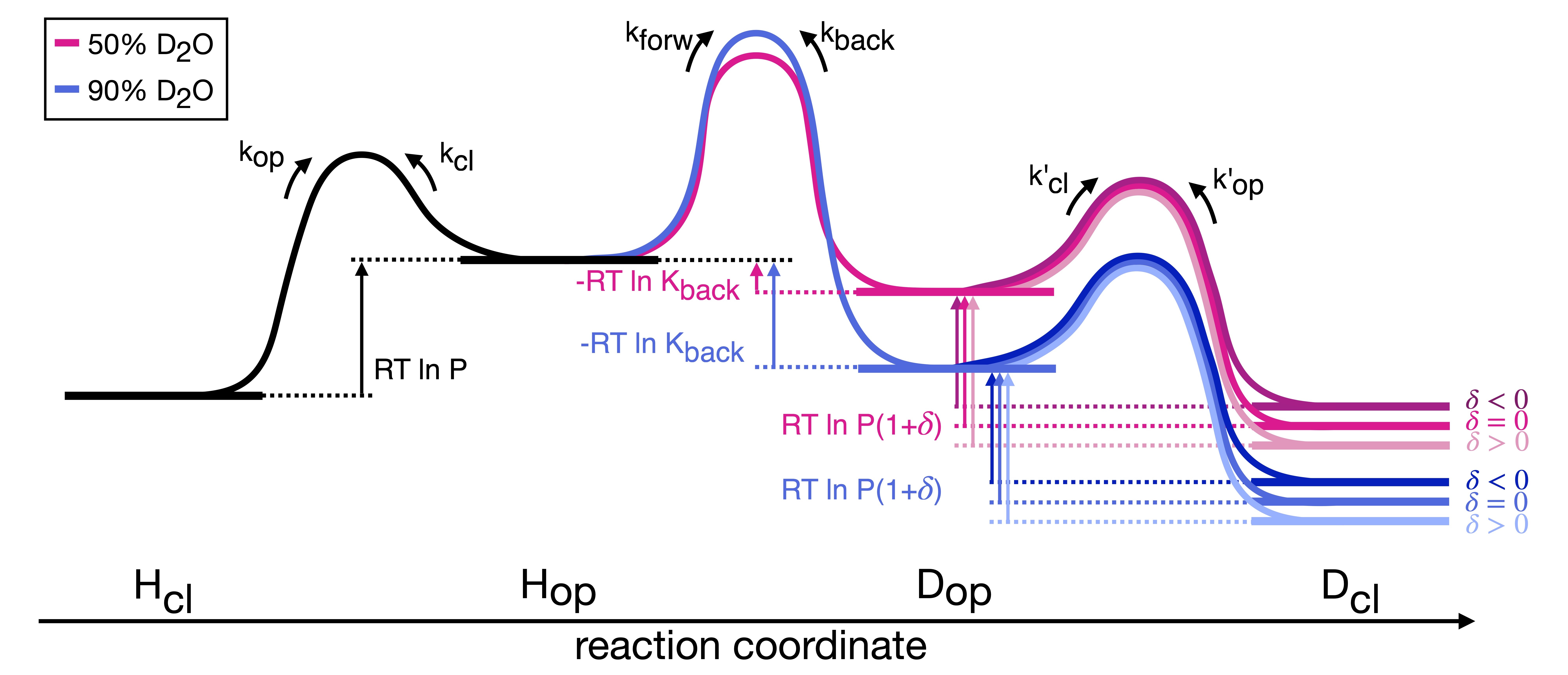}
    \caption{Energy diagram for the GLL model \eqref{eq:gll}. Exchange in the EX2 limit (high protection and exchange slower than opening/closing dynamics) is considered in two hypothetical mixtures at same pH and temperature and different compositions, 50\% (magenta) and 90\% (blue) \ce{D2O}. Rate constants $k^\ddagger$ are related to the height of the corresponding barriers $\Delta G^\ddagger$ by Eyring equation: $k^\ddagger = (k_\mathrm{B}T/h) \mathrm{e}^{-\Delta G^\ddagger/RT}$. The solvent composition affects $k_\mathrm{forw}$, $k_\mathrm{back}$ and their ratio $K_\mathrm{back}$, determining the equilibrium between open states. A change in opening free energy upon isotopic substitution that is quantified by $\delta$, \textit{cfr} Eq.\! \ref{eq:delta-delta-G}, affects equilibrium between deuterated states. The height of the barrier depends on $k_\mathrm{cl}'$ and $k_\mathrm{op}'$. In this illustration, it is assumed $k_\mathrm{cl}'=k_\mathrm{cl}$ and $k_\mathrm{op}'$ varying according to $\delta$.}
    \label{fig:TS}
\end{figure}

Amide HDX in a \ce{H2O/D2O} mixture can be described by the generalized Linderstr{\o}m-Lang (GLL) model
\begin{align}
    \ce{H_{cl}} \xrightleftharpoons[k_\mathrm{cl}]{k_\mathrm{op}} \ce{H_{op}} \xrightleftharpoons[k_\mathrm{back}]{k_\mathrm{forw}} \ce{D_{op}} \xrightleftharpoons[k_\mathrm{op}']{k_\mathrm{cl}'} \ce{D_{cl}}.
\label{eq:gll}
\end{align}
Open and closed states are defined as for the LL model \eqref{eq:ll}, and their interconversion rates differ upon isotopic substitution, i.e.\! $k_\mathrm{cl}'\neq k_\mathrm{cl}$ and $k_\mathrm{op}' \neq k_\mathrm{op}$.
This implies a different protection factor for the deuterated amide $P'=k_\mathrm{cl}'/k_\mathrm{op}'$. $P'$ can be written as $P'=P(1+\delta)$, where $\delta$ can be positive or negative and determines the difference in opening free energies 
\begin{align}
    \Delta \Delta G_\mathrm{op} = \Delta G_\mathrm{op, D} - \Delta G_\mathrm{op,H} = RT \ln (1 + \delta),
\label{eq:delta-delta-G}
\end{align}
where $\Delta G_\mathrm{op,H}$ is defined in Eq.\! \ref{eq:ll-DeltaG=RTlnP} and $\Delta G_\mathrm{op,D} = RT \ln P'$.
For unprotected amides, exchange occurs with forward and back exchange rate constants $k_\mathrm{forw}$ and $k_\mathrm{back}$ that depend on sequence, temperature and pH analogously to $k_\mathrm{int}$ in the LL model, as well as \ce{D2O} content.
Forward and back exchange rate constants can be estimated as described in ref.\! \citenum{grimaldi2025method}.
The approach-to-equilibrium rate of the elementary exchange reaction defines an intrinsic HDX rate in the mixture $k_\mathrm{int,mix} = k_\mathrm{forw} + k_\mathrm{back}$, and the equilibrium constant of the back exchange reaction is $K_\mathrm{back} = k_\mathrm{back}/k_\mathrm{forw}$.
An energy diagram of the GLL model \eqref{eq:gll} is sketched in Figure \ref{fig:TS}.

Rate equations for the GLL model \eqref{eq:gll} can be exactly solved and converge to a stationary state.
Closed-form expressions are obtained in approximations that involve separation of time scales, \textit{cfr} Supporting Information.

The analogous of the EX2 approximation, obtained assuming $P,P' \gg 1$ and $k_\mathrm{cl}, k_\mathrm{cl}' \gg k_\mathrm{int,mix}$, gives
\begin{align}
    D(t) = D_\mathrm{eq} + (D_0 - D_\mathrm{eq}) \mathrm{e}^{-k_\mathrm{obs}t},
\label{eq:gll-single-exp}
\end{align}
where $D_0$ is the initial condition,
\begin{align}
    D_\mathrm{eq} = \frac{1 + \delta}{1 + \delta + K_\mathrm{back}}
\label{eq:gll-ex2-Deq}
\end{align}
is the fraction of deuterated amides at equilibrium, and
\begin{align}
    k_\mathrm{obs} = \frac{k_\mathrm{int,mix}}{P} \left( 1 - \frac{K_\mathrm{back}}{1 + K_\mathrm{back}} \frac{\delta}{1+\delta} \right).
\label{eq:gll-ex2-kobs}
\end{align}

From Eqs.\! \ref{eq:gll-ex2-Deq} and \ref{eq:gll-ex2-kobs}, it results that back exchange and the difference in protection upon deuteration determine equilibrium and kinetics of HDX reactions.
In a mixture with \ce{D2O} mole fraction $x$ (\ce{H2O} mole fraction $1-x$), in general $D_\mathrm{eq} \neq x$.
The equilibrium ratio
\begin{align}
    \frac{D_\mathrm{eq}}{H_\mathrm{eq}} = \frac{1+\delta}{K_\mathrm{back}}
\label{eq:gll-equil-ratio}
\end{align}
is related to the fractionation factor
\begin{align}
    \varphi = \frac{D_\mathrm{eq}}{H_\mathrm{eq}} \frac{1-x}{x}
\label{eq:fractionation-factor}
\end{align}
that quantifies amide enrichment in deuterium with respect to the solvent \cite{gold1969protolytic}.
If $\delta = 0$, the effect on kinetics amounts to replacing $k_\mathrm{int}$ from Eq.\! \ref{eq:ll-ex2-kobs} with $k_\mathrm{int,mix}$ that accounts for simultaneous forward and reverse exchange.
In the broader case $\delta \neq 0$, an additional term that depends on both $\delta$ and $K_\mathrm{back}$ appears, \textit{cfr} Eq.\! \ref{eq:gll-ex2-kobs}.
In pure \ce{D2O}, $k_\mathrm{forw} = k_\mathrm{int}$ and $k_\mathrm{back} = 0$, which imply $k_\mathrm{int,mix} = k_\mathrm{int}$ and $K_\mathrm{back} = 0$, hence $D_\mathrm{eq} = 1$, $k_\mathrm{obs} = k_\mathrm{int}/P$, and the LL model is recovered.

\section*{Results and discussion}

\begin{figure}
    \centering
    \includegraphics[width=.9\linewidth]{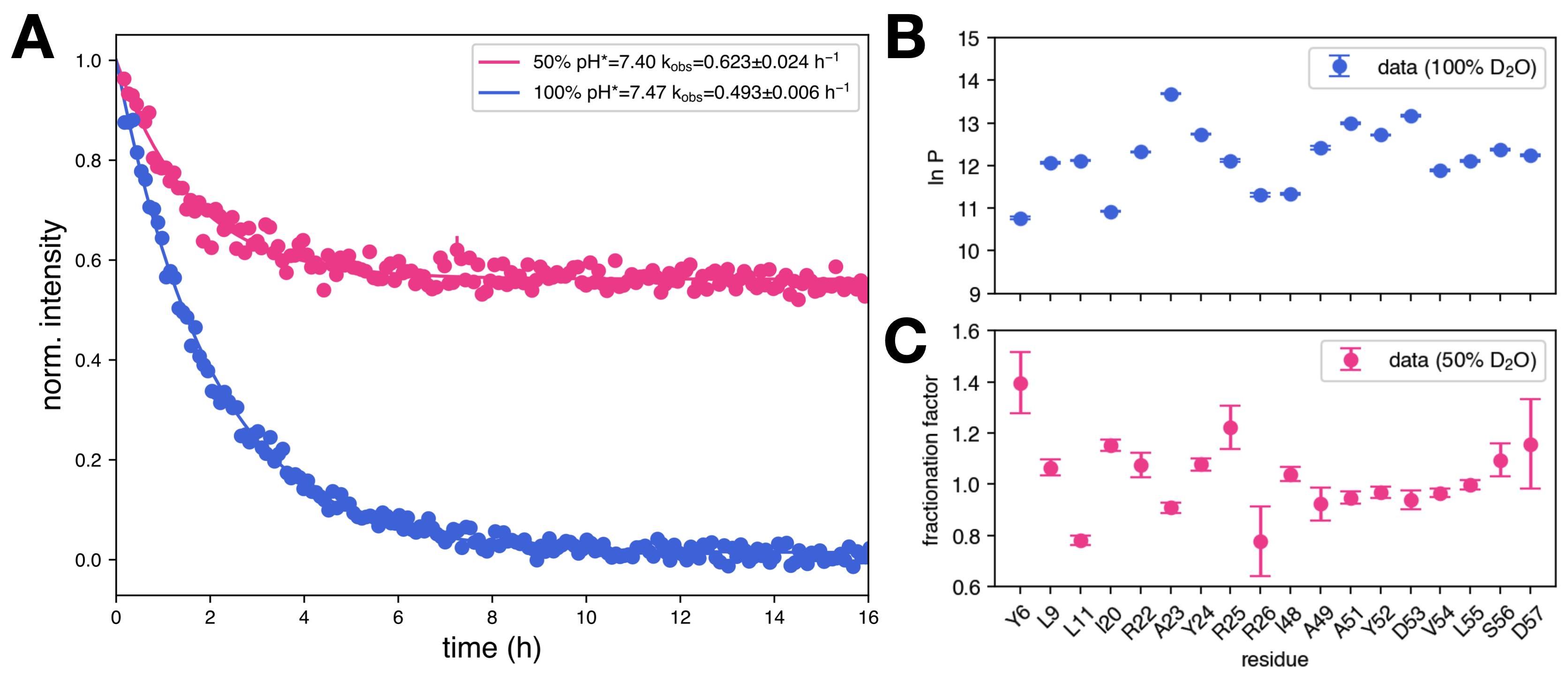}
    \caption{Experimental results for HDX/NMR of \ce{^{15}N}-DNAJB1 JD-GF-${\rm \alpha 5}$ F94L, at 25° C in 50\% \ce{D2O} and $\mathrm{pH_{read}} = 7.40$ and 100\% \ce{D2O} and $\mathrm{pH_{read} = 7.47}$. (A) Measured kinetics of residue L11, in 50\% (magenta) and 100\% (blue) \ce{D2O}. The normalized intensity is the fraction of unexchanged amides, $1-D(t)$. (B) Protection factors ($\ln P$) estimated by measurements in 100\% \ce{D2O}. (C) Fractionation factors determined from equilibrium values in 50\% \ce{D2O}.}
    \label{fig:nmr}
\end{figure}

HDX/NMR measurements of initially undeuterated \ce{^{15}N}-DNAJB1 JD-GF-${\rm \alpha 5}$ F94L \cite{hobbs2025low} were performed at 25° C in 50\% ($\mathrm{pH_{read}}=7.40$) and 100\% ($\mathrm{pH_{read}}=7.47$) \ce{D2O}, \textit{cfr} Methods.
Data points, i.e., peak intensities over time were measured in a series of \ce{^1H}-\ce{^{15}N} SOFAST HMQC spectra and fitted to a single exponential $I(t) = a \mathrm{e}^{-bt} + c$ for 18 residues.
The fraction of unexchanged amides at time $t$ is $1-D(t)=I(t)/I(0)$, where $I(0)=a+c$ is the extrapolated intensity at $t=0$.
The observed exchange rate is $k_\mathrm{obs} = b$.
The equilibrium fractions of deuterated and undeuterated sites are $D_\mathrm{eq} = \frac{a}{a+c}$ and $ H_\mathrm{eq} = \frac{c}{a+c}$, respectively.
Data and fit for residue L11 are shown as an example in Figure \ref{fig:nmr}A.
Curves for all measured amides are provided, \textit{cfr} Supporting Information.
Protection factors were estimated from measurements in 100\% \ce{D2O} using the LL model, that is, by Eq.\! \ref{eq:ll-ex2-kobs} (Figure \ref{fig:nmr}B).
Fractionation factors (Figure \ref{fig:nmr}C) were derived from the parameters of the fit in 50\% \ce{D2O} as $\varphi = D_\mathrm{eq}/H_\mathrm{eq} = a/c$.

\begin{figure}
    \centering
    \includegraphics[width=.45\linewidth]{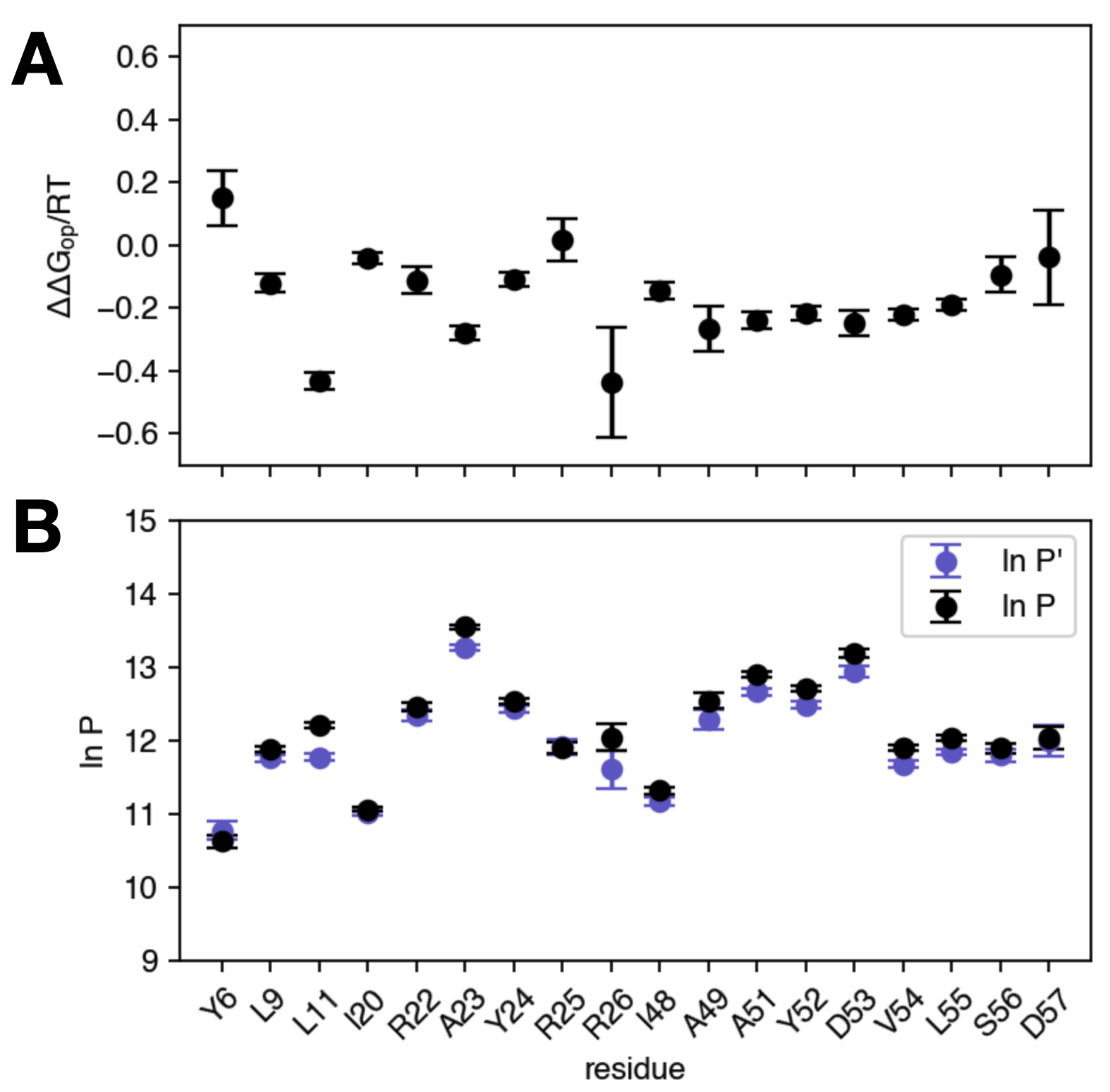}
    \caption{Results obtained by the generalised Linderstr{\o}m-Lang model for HDX/NMR of \ce{^{15}N}-DNAJB1 JD-GF-${\rm \alpha 5}$ F94L, performed in 50\% \ce{D2O}. (A) Difference in local stability (protection factors) upon isotopic substitution quantified by $\Delta\Delta G_\mathrm{op}$. (B) Inferred protection factors for undeuterated ($\ln P$) and deuterated ($\ln P'$) amides.}
    \label{fig:res}
\end{figure}

For measurements performed in mixtures, $k_\mathrm{forw}$ and $k_\mathrm{back}$ were estimated as a function of temperature, pH and sequence, as described in ref.\! \citenum{grimaldi2025method}.
Accordingly, $k_\mathrm{int,mix}$ and $K_\mathrm{back}$ were computed as their sum and their ratio, \textit{vide supra}.
A value of $K_\mathrm{back} = 0.83$ was consistently found for all residues (this because $K_\mathrm{back}$ refers to the exchange of unprotected amides, for which the model\cite{grimaldi2025method} predicts $\varphi = 1.20$, in agreement with reported observations on PDLA\cite{bowers1996hydrogen}).
The parameter $\delta$ was computed from the measured fractionation $\varphi$ as $\delta = K_\mathrm{back} \varphi - 1$ (Eqs.\! \ref{eq:gll-equil-ratio} and \ref{eq:fractionation-factor}), and directly yields the difference in opening free energy $\Delta\Delta G_\mathrm{op}$ resulting from isotopic substitution (Eq.\! \ref{eq:delta-delta-G}), shown in Fig.\! \ref{fig:res}A.
Finally, protection factors were estimated using the GLL model \eqref{eq:gll} in EX2 approximation (Eq.\! \ref{eq:gll-ex2-kobs}), \textit{cfr} Fig.\! \ref{fig:res}B.

\begin{figure}
    \centering
    \includegraphics[width=.9\linewidth]{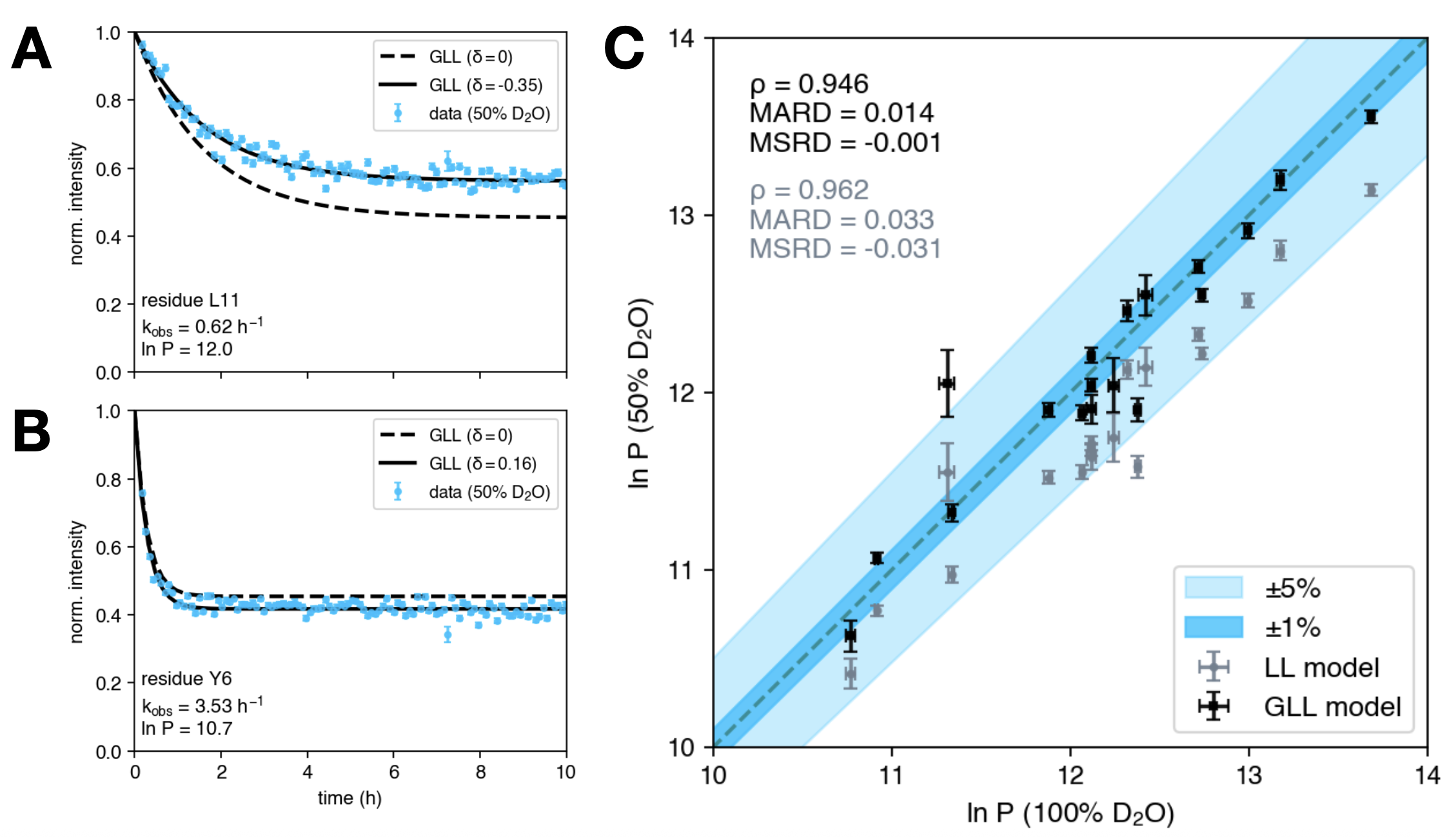}
    \caption{HDX experimental data (dots) obtained from measurements in 50\% \ce{D2O} for residues (A) L11, and (B) Y6, and reproduced by the GLL model. Solid lines are generated from the full model, in which $\delta \neq 0$ indicates variation in local stability upon isotopic substitution and results in fractionation, as well as minor alteration to the kinetics. Dashed lines are obtained considering same $k_\mathrm{int,mix}$ and $P$, and $\delta = 0$. (C) Pair plot of protection factors extracted from data in 100\% \ce{D2O} versus protection factors estimated by LL (gray dots) and GLL (black dots) model from data in 50\% \ce{D2O}. The GLL model results present no systematic bias, while protection factors computed using the LL model are systematically underestimated.}
    \label{fig:val}
\end{figure}

Figures \ref{fig:val}A and \ref{fig:val}B display HDX curves for residues L11 and Y6 fitted by the GLL model. In the former, a reduction in local stability upon isotope substitution ($\delta < 0$) causes an upward shift of the plateau, i.e., favors retention of H isotope. In the latter, a shift in the opposite direction is observed ($\delta > 0$).

The protection factors obtained in 50\% \ce{D2O} using the GLL model can be then compared with the results from pure \ce{D2O}. The two sets are found to be highly correlated (Pearson's $\rho = 0.946$). To quantify their accuracy and bias, one can consider the mean absolute relative deviation (MARD), \[ \mathrm{MARD} = \frac{1}{N} \sum_{i=1}^N |\varepsilon_i|, \] and the mean signed relative deviation (MSRD), \[ \mathrm{MSRD} = \frac{1}{N} \sum_{i=1}^N \varepsilon_i, \]
where \[ \varepsilon_i = \frac{ \ln P_{i,50\%} - \ln P_{i,100\%} }{ \ln P_{i,100\%} } \] is the relative deviation between protection factors inferred in 50\% and 100\% \ce{D2O} for the $i$-th measured residue ($i=1,2,\dots,N=18$).
An analogous comparison was made with the protection factors one would recover using the LL formula (Eq.\! \ref{eq:ll-ex2-kobs}), as a function of the observed exchange rate only, and considering the intrinsic rate in pure \ce{D2O}.
Results from both methods are shown in Figure \ref{fig:val}C.
Accounting for back exchange through the GLL model improves accuracy in protection factors prediction, as testified by the lower MARD (0.014 vs.\! 0.033).
Moreover, the very low MSRD (-0.001) indicates that the GLL model is unbiased, while the more substantial MSRD (-0.031) resulting from the LL model suggests a systematic underestimation of protection factors.
The relative error between protection factors estimated in 50\% \ce{D2O} using the GLL and those in pure \ce{D2O} is below 1\% for half of the probed amides and below 5\% for all the remaining, except R26, which appears to be overestimated by both LL and GLL model
(a possible source of error is the inadequate sampling of the curve in 100\% \ce{D2O} that lead to an incorrect estimation of $P$).
The Linderstrøm-Lang model provides straightforwardly (from Eq. \ref{eq:ll-ex2-kobs}) protection factors from exchange measurement in 100\% \ce{D2O} (Fig.\! \ref{fig:nmr}B).
The GLL model presented here, remarkably,  simultaneously provides two parameters: protection factors and fractionation factors from a single HDX experiment in a mixture (e.g., 50\% \ce{D2O}).
This dual information content is not accessible from classical experiments in pure \ce{D2O} which yield protection factors alone. Fractionation factors thus constitute an additional structural observable that can be used to refine models of protein conformational ensembles and provide complementary insights into hydrogen-bonding networks \cite{cleland1994low, loh1994hydrogen, edison1995theoretical, bowers1996hydrogen, krantz2000d}. 




\section*{Methods}

Samples of 15N-DNAJB1 JD-GF-${\rm \alpha 5}$ F94L were expressed and purified as described previously \cite{hobbs2025low}. In particular, samples  were prepared in 20 mM sodium phosphate pH 7.0, 50 mM \ce{NaCl} and lyophilised. Freeze-dried protein was resuspended in 50 and 100\% (v/v) \ce{D2O}, placed into an NMR tube and the loss of intensity of amide protons was monitored using SOFAST $^{1}\ce{H}$-$^{15}\ce{N}$ HMQC experiments at 25° C and 600 MHz. Experiments were recorded with 128 increments in the indirect dimension with 4 scans per increment, 1024 complex points and a D1 of 0.5 s for a total experimental time of about 5 minutes.

\begin{acknowledgement}
AG acknowledges a PhD studentship funded by PNRR. Solution NMR studies were supported by a Sir Henry Dale Fellowship jointly funded by the Wellcome Trust and the Royal Society (Grant Number 223268/Z/21/Z) to TKK.
\end{acknowledgement}

\bibliography{bib_main}

\providecommand{\noopsort}[1]{}\providecommand{\singleletter}[1]{#1}%
\providecommand{\latin}[1]{#1}
\makeatletter
\providecommand{\doi}
  {\begingroup\let\do\@makeother\dospecials
  \catcode`\{=1 \catcode`\}=2 \doi@aux}
\providecommand{\doi@aux}[1]{\endgroup\texttt{#1}}
\makeatother
\providecommand*\mcitethebibliography{\thebibliography}
\csname @ifundefined\endcsname{endmcitethebibliography}  {\let\endmcitethebibliography\endthebibliography}{}
\begin{mcitethebibliography}{29}
\providecommand*\natexlab[1]{#1}
\providecommand*\mciteSetBstSublistMode[1]{}
\providecommand*\mciteSetBstMaxWidthForm[2]{}
\providecommand*\mciteBstWouldAddEndPuncttrue
  {\def\EndOfBibitem{\unskip.}}
\providecommand*\mciteBstWouldAddEndPunctfalse
  {\let\EndOfBibitem\relax}
\providecommand*\mciteSetBstMidEndSepPunct[3]{}
\providecommand*\mciteSetBstSublistLabelBeginEnd[3]{}
\providecommand*\EndOfBibitem{}
\mciteSetBstSublistMode{f}
\mciteSetBstMaxWidthForm{subitem}{(\alph{mcitesubitemcount})}
\mciteSetBstSublistLabelBeginEnd
  {\mcitemaxwidthsubitemform\space}
  {\relax}
  {\relax}

\bibitem[Hvidt and Linderstr{\o}m-Lang(1954)Hvidt, and Linderstr{\o}m-Lang]{hvidt1954exchange}
Hvidt,~A.; Linderstr{\o}m-Lang,~K. Exchange of hydrogen atoms in insulin with deuterium atoms in aqueous solutions. \emph{Biochimica et biophysica acta} \textbf{1954}, \emph{14}, 574--575\relax
\mciteBstWouldAddEndPuncttrue
\mciteSetBstMidEndSepPunct{\mcitedefaultmidpunct}
{\mcitedefaultendpunct}{\mcitedefaultseppunct}\relax
\EndOfBibitem
\bibitem[Linderstr{\o}m-Lang(1955)]{linderstrom1955ph}
Linderstr{\o}m-Lang,~K. The {pH}-dependence of the deuterium exchange of insulin. \emph{Biochimica et biophysica acta} \textbf{1955}, \emph{18}, 308\relax
\mciteBstWouldAddEndPuncttrue
\mciteSetBstMidEndSepPunct{\mcitedefaultmidpunct}
{\mcitedefaultendpunct}{\mcitedefaultseppunct}\relax
\EndOfBibitem
\bibitem[Englander \latin{et~al.}(1997)Englander, Mayne, Bai, and Sosnick]{englander1997hydrogen}
Englander,~S.; Mayne,~L.; Bai,~Y.; Sosnick,~T. Hydrogen exchange: the modern legacy of Linderstr{\o}m-Lang. \emph{Protein science} \textbf{1997}, \emph{6}, 1101--1109\relax
\mciteBstWouldAddEndPuncttrue
\mciteSetBstMidEndSepPunct{\mcitedefaultmidpunct}
{\mcitedefaultendpunct}{\mcitedefaultseppunct}\relax
\EndOfBibitem
\bibitem[Hvidt and Nielsen(1966)Hvidt, and Nielsen]{hvidt1966hydrogen}
Hvidt,~A.; Nielsen,~S.~O. Hydrogen exchange in proteins. \emph{Advances in protein chemistry} \textbf{1966}, \emph{21}, 287--386\relax
\mciteBstWouldAddEndPuncttrue
\mciteSetBstMidEndSepPunct{\mcitedefaultmidpunct}
{\mcitedefaultendpunct}{\mcitedefaultseppunct}\relax
\EndOfBibitem
\bibitem[Englander and Kallenbach(1983)Englander, and Kallenbach]{englander1983hydrogen}
Englander,~S.~W.; Kallenbach,~N.~R. Hydrogen exchange and structural dynamics of proteins and nucleic acids. \emph{Quarterly reviews of biophysics} \textbf{1983}, \emph{16}, 521--655\relax
\mciteBstWouldAddEndPuncttrue
\mciteSetBstMidEndSepPunct{\mcitedefaultmidpunct}
{\mcitedefaultendpunct}{\mcitedefaultseppunct}\relax
\EndOfBibitem
\bibitem[Hamuro(2020)]{hamuro2020tutorial}
Hamuro,~Y. Tutorial: chemistry of hydrogen/deuterium exchange mass spectrometry. \emph{Journal of the American Society for Mass Spectrometry} \textbf{2020}, \emph{32}, 133--151\relax
\mciteBstWouldAddEndPuncttrue
\mciteSetBstMidEndSepPunct{\mcitedefaultmidpunct}
{\mcitedefaultendpunct}{\mcitedefaultseppunct}\relax
\EndOfBibitem
\bibitem[James \latin{et~al.}(2021)James, Murphree, Vorauer, Engen, and Guttman]{james2021advances}
James,~E.~I.; Murphree,~T.~A.; Vorauer,~C.; Engen,~J.~R.; Guttman,~M. Advances in hydrogen/deuterium exchange mass spectrometry and the pursuit of challenging biological systems. \emph{Chemical reviews} \textbf{2021}, \emph{122}, 7562--7623\relax
\mciteBstWouldAddEndPuncttrue
\mciteSetBstMidEndSepPunct{\mcitedefaultmidpunct}
{\mcitedefaultendpunct}{\mcitedefaultseppunct}\relax
\EndOfBibitem
\bibitem[Bai \latin{et~al.}(1993)Bai, Milne, Mayne, and Englander]{bai1993primary}
Bai,~Y.; Milne,~J.~S.; Mayne,~L.; Englander,~S.~W. Primary structure effects on peptide group hydrogen exchange. \emph{Proteins: Structure, Function, and Bioinformatics} \textbf{1993}, \emph{17}, 75--86\relax
\mciteBstWouldAddEndPuncttrue
\mciteSetBstMidEndSepPunct{\mcitedefaultmidpunct}
{\mcitedefaultendpunct}{\mcitedefaultseppunct}\relax
\EndOfBibitem
\bibitem[Connelly \latin{et~al.}(1993)Connelly, Bai, Jeng, and Englander]{connelly1993isotope}
Connelly,~G.~P.; Bai,~Y.; Jeng,~M.-F.; Englander,~S.~W. Isotope effects in peptide group hydrogen exchange. \emph{Proteins: Structure, Function, and Bioinformatics} \textbf{1993}, \emph{17}, 87--92\relax
\mciteBstWouldAddEndPuncttrue
\mciteSetBstMidEndSepPunct{\mcitedefaultmidpunct}
{\mcitedefaultendpunct}{\mcitedefaultseppunct}\relax
\EndOfBibitem
\bibitem[Nguyen \latin{et~al.}(2018)Nguyen, Mayne, Phillips, and Walter~Englander]{nguyen2018reference}
Nguyen,~D.; Mayne,~L.; Phillips,~M.~C.; Walter~Englander,~S. Reference parameters for protein hydrogen exchange rates. \emph{Journal of the American Society for Mass Spectrometry} \textbf{2018}, \emph{29}, 1936--1939\relax
\mciteBstWouldAddEndPuncttrue
\mciteSetBstMidEndSepPunct{\mcitedefaultmidpunct}
{\mcitedefaultendpunct}{\mcitedefaultseppunct}\relax
\EndOfBibitem
\bibitem[Englander \latin{et~al.}(2016)Englander, Mayne, Kan, and Hu]{englander2016protein}
Englander,~S.~W.; Mayne,~L.; Kan,~Z.-Y.; Hu,~W. Protein folding—how and why: by hydrogen exchange, fragment separation, and mass spectrometry. \emph{Annual Review of Biophysics} \textbf{2016}, \emph{45}, 135--152\relax
\mciteBstWouldAddEndPuncttrue
\mciteSetBstMidEndSepPunct{\mcitedefaultmidpunct}
{\mcitedefaultendpunct}{\mcitedefaultseppunct}\relax
\EndOfBibitem
\bibitem[Englander(2023)]{englander2023hx}
Englander,~S.~W. HX and Me: Understanding allostery, folding, and protein machines. \emph{Annual Review of Biophysics} \textbf{2023}, \emph{52}, 1--18\relax
\mciteBstWouldAddEndPuncttrue
\mciteSetBstMidEndSepPunct{\mcitedefaultmidpunct}
{\mcitedefaultendpunct}{\mcitedefaultseppunct}\relax
\EndOfBibitem
\bibitem[Masson \latin{et~al.}(2017)Masson, Jenkins, and Burke]{masson2017overview}
Masson,~G.~R.; Jenkins,~M.~L.; Burke,~J.~E. An overview of hydrogen deuterium exchange mass spectrometry (HDX-MS) in drug discovery. \emph{Expert opinion on drug discovery} \textbf{2017}, \emph{12}, 981--994\relax
\mciteBstWouldAddEndPuncttrue
\mciteSetBstMidEndSepPunct{\mcitedefaultmidpunct}
{\mcitedefaultendpunct}{\mcitedefaultseppunct}\relax
\EndOfBibitem
\bibitem[Krishnamurthy \latin{et~al.}(2025)Krishnamurthy, Musgaard, Tehan, Jazayeri, and Liko]{krishnamurthy2025evolving}
Krishnamurthy,~S.; Musgaard,~M.; Tehan,~B.~G.; Jazayeri,~A.; Liko,~I. The evolving role of hydrogen/deuterium exchange mass spectrometry in early-stage drug discovery. \emph{Current Opinion in Structural Biology} \textbf{2025}, \emph{92}, 103051\relax
\mciteBstWouldAddEndPuncttrue
\mciteSetBstMidEndSepPunct{\mcitedefaultmidpunct}
{\mcitedefaultendpunct}{\mcitedefaultseppunct}\relax
\EndOfBibitem
\bibitem[Devaurs \latin{et~al.}(2022)Devaurs, Antunes, and Borysik]{devaurs2022computational}
Devaurs,~D.; Antunes,~D.~A.; Borysik,~A.~J. Computational modeling of molecular structures guided by hydrogen-exchange data. \emph{Journal of the American Society for Mass Spectrometry} \textbf{2022}, \emph{33}, 215--237\relax
\mciteBstWouldAddEndPuncttrue
\mciteSetBstMidEndSepPunct{\mcitedefaultmidpunct}
{\mcitedefaultendpunct}{\mcitedefaultseppunct}\relax
\EndOfBibitem
\bibitem[Ghafouri \latin{et~al.}(2025)Ghafouri, Kade{\v{r}}{\'a}vek, Melo, Aspromonte, Bernad{\'o}, Cortes, Doszt{\'a}nyi, Erdos, Feig, Janson, \latin{et~al.} others]{ghafouri2025unifiedframeworkdeterminingconformational}
Ghafouri,~H.; Kade{\v{r}}{\'a}vek,~P.; Melo,~A.~M.; Aspromonte,~M.~C.; Bernad{\'o},~P.; Cortes,~J.; Doszt{\'a}nyi,~Z.; Erdos,~G.; Feig,~M.; Janson,~G.; others Towards a Unified Framework for Determining Conformational Ensembles of Disordered Proteins. \emph{arXiv preprint arXiv:2504.03590} \textbf{2025}, \relax
\mciteBstWouldAddEndPunctfalse
\mciteSetBstMidEndSepPunct{\mcitedefaultmidpunct}
{}{\mcitedefaultseppunct}\relax
\EndOfBibitem
\bibitem[Engen and Wales(2015)Engen, and Wales]{engen2015analytical}
Engen,~J.~R.; Wales,~T.~E. Analytical aspects of hydrogen exchange mass spectrometry. \emph{Annual Review of Analytical Chemistry} \textbf{2015}, \emph{8}, 127--148\relax
\mciteBstWouldAddEndPuncttrue
\mciteSetBstMidEndSepPunct{\mcitedefaultmidpunct}
{\mcitedefaultendpunct}{\mcitedefaultseppunct}\relax
\EndOfBibitem
\bibitem[Stofella \latin{et~al.}(2024)Stofella, Grimaldi, Smit, Claesen, Paci, and Sobott]{stofella2024computational}
Stofella,~M.; Grimaldi,~A.; Smit,~J.~H.; Claesen,~J.; Paci,~E.; Sobott,~F. Computational Tools for Hydrogen--Deuterium Exchange Mass Spectrometry Data Analysis. \emph{Chemical Reviews} \textbf{2024}, \emph{124}, 12242--12263\relax
\mciteBstWouldAddEndPuncttrue
\mciteSetBstMidEndSepPunct{\mcitedefaultmidpunct}
{\mcitedefaultendpunct}{\mcitedefaultseppunct}\relax
\EndOfBibitem
\bibitem[Konermann and Scrosati(2024)Konermann, and Scrosati]{konermann-opportunities}
Konermann,~L.; Scrosati,~P.~M. Hydrogen/Deuterium Exchange Mass Spectrometry: Fundamentals, Limitations, and Opportunities. \emph{Molecular \& Cellular Proteomics: MCP} \textbf{2024}, \emph{23}, 100853\relax
\mciteBstWouldAddEndPuncttrue
\mciteSetBstMidEndSepPunct{\mcitedefaultmidpunct}
{\mcitedefaultendpunct}{\mcitedefaultseppunct}\relax
\EndOfBibitem
\bibitem[Cleland and Kreevoy(1994)Cleland, and Kreevoy]{cleland1994low}
Cleland,~W.; Kreevoy,~M.~M. Low-barrier hydrogen bonds and enzymic catalysis. \emph{Science} \textbf{1994}, \emph{264}, 1887--1890\relax
\mciteBstWouldAddEndPuncttrue
\mciteSetBstMidEndSepPunct{\mcitedefaultmidpunct}
{\mcitedefaultendpunct}{\mcitedefaultseppunct}\relax
\EndOfBibitem
\bibitem[Loh and Markley(1994)Loh, and Markley]{loh1994hydrogen}
Loh,~S.~N.; Markley,~J.~L. Hydrogen bonding in proteins as studied by amide hydrogen D/H fractionation factors: application to staphylococcal nuclease. \emph{Biochemistry} \textbf{1994}, \emph{33}, 1029--1036\relax
\mciteBstWouldAddEndPuncttrue
\mciteSetBstMidEndSepPunct{\mcitedefaultmidpunct}
{\mcitedefaultendpunct}{\mcitedefaultseppunct}\relax
\EndOfBibitem
\bibitem[Edison \latin{et~al.}(1995)Edison, Weinhold, and Markley]{edison1995theoretical}
Edison,~A.~S.; Weinhold,~F.; Markley,~J.~L. Theoretical studies of protium/deuterium fractionation factors and cooperative hydrogen bonding in peptides. \emph{Journal of the American Chemical Society} \textbf{1995}, \emph{117}, 9619--9624\relax
\mciteBstWouldAddEndPuncttrue
\mciteSetBstMidEndSepPunct{\mcitedefaultmidpunct}
{\mcitedefaultendpunct}{\mcitedefaultseppunct}\relax
\EndOfBibitem
\bibitem[Bowers and Klevit(1996)Bowers, and Klevit]{bowers1996hydrogen}
Bowers,~P.~M.; Klevit,~R.~E. Hydrogen bonding and equilibrium isotope enrichment in histidine-containing proteins. \emph{Nature Structural Biology} \textbf{1996}, \emph{3}, 522--531\relax
\mciteBstWouldAddEndPuncttrue
\mciteSetBstMidEndSepPunct{\mcitedefaultmidpunct}
{\mcitedefaultendpunct}{\mcitedefaultseppunct}\relax
\EndOfBibitem
\bibitem[LiWang and Bax(1996)LiWang, and Bax]{liwang1996equilibrium}
LiWang,~A.~C.; Bax,~A. Equilibrium protium/deuterium fractionation of backbone amides in U- 13C/15N labeled human ubiquitin by triple resonance NMR. \emph{Journal of the American Chemical Society} \textbf{1996}, \emph{118}, 12864--12865\relax
\mciteBstWouldAddEndPuncttrue
\mciteSetBstMidEndSepPunct{\mcitedefaultmidpunct}
{\mcitedefaultendpunct}{\mcitedefaultseppunct}\relax
\EndOfBibitem
\bibitem[Krantz \latin{et~al.}(2000)Krantz, Moran, Kentsis, and Sosnick]{krantz2000d}
Krantz,~B.~A.; Moran,~L.~B.; Kentsis,~A.; Sosnick,~T.~R. D/H amide kinetic isotope effects reveal when hydrogen bonds form during protein folding. \emph{nature structural biology} \textbf{2000}, \emph{7}, 62--71\relax
\mciteBstWouldAddEndPuncttrue
\mciteSetBstMidEndSepPunct{\mcitedefaultmidpunct}
{\mcitedefaultendpunct}{\mcitedefaultseppunct}\relax
\EndOfBibitem
\bibitem[Grimaldi \latin{et~al.}(2025)Grimaldi, Stofella, and Paci]{grimaldi2025method}
Grimaldi,~A.; Stofella,~M.; Paci,~E. Method to estimate amide base-catalyzed back exchange rates in H2O/D2O mixtures. \emph{bioRxiv} \textbf{2025}, 2025--09\relax
\mciteBstWouldAddEndPuncttrue
\mciteSetBstMidEndSepPunct{\mcitedefaultmidpunct}
{\mcitedefaultendpunct}{\mcitedefaultseppunct}\relax
\EndOfBibitem
\bibitem[Gold(1969)]{gold1969protolytic}
Gold,~V. \emph{Advances in Physical Organic Chemistry}; Elsevier, 1969; Vol.~7; pp 259--331\relax
\mciteBstWouldAddEndPuncttrue
\mciteSetBstMidEndSepPunct{\mcitedefaultmidpunct}
{\mcitedefaultendpunct}{\mcitedefaultseppunct}\relax
\EndOfBibitem
\bibitem[Hobbs \latin{et~al.}(2025)Hobbs, Limmer, Ossa, Kn{\"u}pling, Lenton, Foder{\`a}, Kalverda, and Karamanos]{hobbs2025low}
Hobbs,~B.; Limmer,~N.; Ossa,~F.; Kn{\"u}pling,~E.; Lenton,~S.; Foder{\`a},~V.; Kalverda,~A.~P.; Karamanos,~T.~K. A low-complexity linker as a driver of intra-and intermolecular interactions in DNAJB chaperones. \emph{Nature Communications} \textbf{2025}, \emph{16}, 1--14\relax
\mciteBstWouldAddEndPuncttrue
\mciteSetBstMidEndSepPunct{\mcitedefaultmidpunct}
{\mcitedefaultendpunct}{\mcitedefaultseppunct}\relax
\EndOfBibitem
\end{mcitethebibliography}


\providecommand{\noopsort}[1]{}\providecommand{\singleletter}[1]{#1}%
\providecommand{\latin}[1]{#1}
\makeatletter
\providecommand{\doi}
  {\begingroup\let\do\@makeother\dospecials
  \catcode`\{=1 \catcode`\}=2 \doi@aux}
\providecommand{\doi@aux}[1]{\endgroup\texttt{#1}}
\makeatother
\providecommand*\mcitethebibliography{\thebibliography}
\csname @ifundefined\endcsname{endmcitethebibliography}  {\let\endmcitethebibliography\endthebibliography}{}
\begin{mcitethebibliography}{8}
\providecommand*\natexlab[1]{#1}
\providecommand*\mciteSetBstSublistMode[1]{}
\providecommand*\mciteSetBstMaxWidthForm[2]{}
\providecommand*\mciteBstWouldAddEndPuncttrue
  {\def\EndOfBibitem{\unskip.}}
\providecommand*\mciteBstWouldAddEndPunctfalse
  {\let\EndOfBibitem\relax}
\providecommand*\mciteSetBstMidEndSepPunct[3]{}
\providecommand*\mciteSetBstSublistLabelBeginEnd[3]{}
\providecommand*\EndOfBibitem{}
\mciteSetBstSublistMode{f}
\mciteSetBstMaxWidthForm{subitem}{(\alph{mcitesubitemcount})}
\mciteSetBstSublistLabelBeginEnd
  {\mcitemaxwidthsubitemform\space}
  {\relax}
  {\relax}

\bibitem[Hvidt and Nielsen(1966)Hvidt, and Nielsen]{hvidt1966hydrogen}
Hvidt,~A.; Nielsen,~S.~O. Hydrogen exchange in proteins. \emph{Advances in protein chemistry} \textbf{1966}, \emph{21}, 287--386\relax
\mciteBstWouldAddEndPuncttrue
\mciteSetBstMidEndSepPunct{\mcitedefaultmidpunct}
{\mcitedefaultendpunct}{\mcitedefaultseppunct}\relax
\EndOfBibitem
\bibitem[Hilser and Freire(1996)Hilser, and Freire]{hilser1996structure}
Hilser,~V.~J.; Freire,~E. Structure-based calculation of the equilibrium folding pathway of proteins. Correlation with hydrogen exchange protection factors. \emph{Journal of molecular biology} \textbf{1996}, \emph{262}, 756--772\relax
\mciteBstWouldAddEndPuncttrue
\mciteSetBstMidEndSepPunct{\mcitedefaultmidpunct}
{\mcitedefaultendpunct}{\mcitedefaultseppunct}\relax
\EndOfBibitem
\bibitem[Bai \latin{et~al.}(1993)Bai, Milne, Mayne, and Englander]{bai1993primary}
Bai,~Y.; Milne,~J.~S.; Mayne,~L.; Englander,~S.~W. Primary structure effects on peptide group hydrogen exchange. \emph{Proteins: Structure, Function, and Bioinformatics} \textbf{1993}, \emph{17}, 75--86\relax
\mciteBstWouldAddEndPuncttrue
\mciteSetBstMidEndSepPunct{\mcitedefaultmidpunct}
{\mcitedefaultendpunct}{\mcitedefaultseppunct}\relax
\EndOfBibitem
\bibitem[Connelly \latin{et~al.}(1993)Connelly, Bai, Jeng, and Englander]{connelly1993isotope}
Connelly,~G.~P.; Bai,~Y.; Jeng,~M.-F.; Englander,~S.~W. Isotope effects in peptide group hydrogen exchange. \emph{Proteins: Structure, Function, and Bioinformatics} \textbf{1993}, \emph{17}, 87--92\relax
\mciteBstWouldAddEndPuncttrue
\mciteSetBstMidEndSepPunct{\mcitedefaultmidpunct}
{\mcitedefaultendpunct}{\mcitedefaultseppunct}\relax
\EndOfBibitem
\bibitem[Nguyen \latin{et~al.}(2018)Nguyen, Mayne, Phillips, and Walter~Englander]{nguyen2018reference}
Nguyen,~D.; Mayne,~L.; Phillips,~M.~C.; Walter~Englander,~S. Reference parameters for protein hydrogen exchange rates. \emph{Journal of the American Society for Mass Spectrometry} \textbf{2018}, \emph{29}, 1936--1939\relax
\mciteBstWouldAddEndPuncttrue
\mciteSetBstMidEndSepPunct{\mcitedefaultmidpunct}
{\mcitedefaultendpunct}{\mcitedefaultseppunct}\relax
\EndOfBibitem
\bibitem[Parlett(1998)]{parlett1998symmetric}
Parlett,~B.~N. \emph{The symmetric eigenvalue problem}; SIAM, 1998\relax
\mciteBstWouldAddEndPuncttrue
\mciteSetBstMidEndSepPunct{\mcitedefaultmidpunct}
{\mcitedefaultendpunct}{\mcitedefaultseppunct}\relax
\EndOfBibitem
\bibitem[Kurzynski(2006)]{kurzynski2006thermodynamic}
Kurzynski,~M. \emph{The Thermodynamic Machinery of Life}; Springer Berlin Heidelberg, 2006\relax
\mciteBstWouldAddEndPuncttrue
\mciteSetBstMidEndSepPunct{\mcitedefaultmidpunct}
{\mcitedefaultendpunct}{\mcitedefaultseppunct}\relax
\EndOfBibitem
\end{mcitethebibliography}

\clearpage

\end{document}


\setstretch{1}

\setcounter{page}{1}
\renewcommand{\thepage}{S\arabic{page}}

\setcounter{equation}{0}
\renewcommand{\theequation}{S\arabic{equation}}

\setcounter{figure}{0}
\renewcommand{\thefigure}{S\arabic{figure}}

\tableofcontents



\newpage

\section{Linderstr{\o}m-Lang model}

The Linderstr{\o}m-Lang (LL) model \cite{hvidt1966hydrogen} describes hydrogen deuterium exchange of proteins in pure \ce{D2O}.
It assumes that each protein backbone amide hydrogen adopts either closed (\ce{H_{cl}}) or open (\ce{H_{op}}) conformations, only the latter being competent to exchange according to the reaction
\begin{align}
    \ce{H_{cl}} \xrightleftharpoons[k_\mathrm{cl}]{k_\mathrm{op}} \ce{H_{op}} \xrightarrow{k_\mathrm{int}} \ce{exchanged}.
\label{eq:si-ll}
\end{align}
The rate constants $k_\mathrm{op}$ and $k_\mathrm{cl}$ that are related to opening and closing transitions encode protein dynamics.
Their ratio $P = k_\mathrm{cl}/k_\mathrm{op}$, which is the reciprocal of the opening equilibrium constant, is called the protection factor and is a key quantity to characterize conformational ensembles \cite{hilser1996structure}.
The protection factor is related to the opening free energy $\Delta G_\mathrm{op} = G_\mathrm{op} - G_\mathrm{cl}$ by
\begin{align}
    \Delta G_\mathrm{op} = RT \ln P.
\label{eq:si-DeltaG=RTlnP}
\end{align}
The intrinsic rate $k_\mathrm{int}$ is estimated as a function of primary structure, temperature, pH \cite{bai1993primary,connelly1993isotope,nguyen2018reference}.

\subsection{Exact solution}

The reaction \eqref{eq:si-ll} is associated to a set of coupled linear differential equations:
\begin{subequations}
\begin{align}
    \dot{H_\mathrm{cl}}(t) &=  -k_\mathrm{op} H_\mathrm{cl}(t) + k_\mathrm{cl} H_\mathrm{op}(t), \label{eq:si-ll-diff-Hcl} \\
    \dot{H_\mathrm{op}}(t) &=  +k_\mathrm{op} H_\mathrm{cl}(t) - ( k_\mathrm{cl} + k_\mathrm{int} ) H_\mathrm{op}(t), \label{eq:si-ll-diff-Hop} \\
    \dot{D}(t) &=  + k_\mathrm{op} + k_\mathrm{int} H_\mathrm{op}(t), \label{eq:si-ll-diff-D}
\end{align}
\end{subequations}
where $H_\mathrm{cl}(t)$, $H_\mathrm{op}(t)$, $D(t)$ are normalized populations, i.e.\! their sum is 1.
Eqs.\! \ref{eq:si-ll-diff-Hcl}, \ref{eq:si-ll-diff-Hop}, \ref{eq:si-ll-diff-D} can be compactly written as $\dot{\mathbf{x}}(t) = \mathbf{Kx}(t)$, where $\mathbf{x}(t) = \begin{bmatrix}
    H_\mathrm{cl}(t) & H_\mathrm{op}(t) & D(t)
\end{bmatrix}^\text{T}$, and
\begin{align}
    \mathbf{K} = \begin{bmatrix}
        -k_\mathrm{op} & +k_\mathrm{cl} & 0 \\
        +k_\mathrm{op} & -k_\mathrm{cl}-k_\mathrm{int} & 0 \\ 
        0 & +k_\mathrm{int} & 0
    \end{bmatrix}.
\label{eq:si-ll-K}
\end{align}
The solution can be written as \[ \mathbf{x}(t) = \sum_{\lambda} c_\lambda \mathrm{e}^{-\lambda t} \mathbf{v}_\lambda,\] where $\{\lambda\}$ and $\{ \mathbf{v}_\lambda \}$ are eigenvalues and eigenvectors of $\mathbf{K}$, while coefficients $\{c_\lambda\}$ depend upon initial conditions.
The eigenvectors and eigenvalues are
\begin{subequations}\label{eq:eig-delta0}
\begin{align}
    &\begin{cases}
        \begin{split}
            \lambda_0 &= 0\\
            \mathbf{v}_0 &= \begin{bmatrix}
                0 & 0 & 1
            \end{bmatrix}^\text{T}
        \end{split}
    \end{cases}\;,
    \label{eq:si-ll-lamb0}\\
    &\begin{cases}
        \begin{split}
            \lambda_\pm &=  - \frac{1}{2} \left( (k_\mathrm{op}+k_\mathrm{cl}+k_\mathrm{int}) \pm  \sqrt{(k_\mathrm{op} + k_\mathrm{cl} + k_\mathrm{int})^2 - 4 k_\mathrm{op} k_\mathrm{int}} \right)\\
            \mathbf{v}_\pm &= \begin{bmatrix}
                \frac{\lambda_\pm}{k_\mathrm{int}} - 1 & -\frac{\lambda_\pm}{k_\mathrm{int}} & 1
            \end{bmatrix}^\text{T}
        \end{split}
    \end{cases}\;.
    \label{eq:si-ll-lambpm}
\end{align}
\end{subequations}

\subsection{Approximations}

The exchanged fraction $D(t)$ is commonly written as a single exponential in the native approximation ($k_\mathrm{cl} \gg k_\mathrm{op}$): \[D(t) = 1 - \mathrm{e}^{-k_\mathrm{obs}t},\] where
\begin{align}
    k_\mathrm{obs} = \frac{k_\mathrm{op} k_\mathrm{int}}{k_\mathrm{cl} + k_\mathrm{int}}
\label{eq:si-ll-native-kobs}
\end{align}
is the observed exchange rate.
This expression suggests two limiting cases depending on the relative magnitude of $k_\mathrm{cl}, k_\mathrm{int}$:
\begin{subequations}\label{eq:ll-ex}
\begin{align}
        \text{EX1}\quad k_\mathrm{cl} \ll k_\mathrm{int} \qquad &k_\mathrm{obs} = k_\mathrm{op}, 
        \label{eq:si-ll-ex1}\\
        \text{EX2}\quad k_\mathrm{cl} \gg k_\mathrm{int} \qquad &k_\mathrm{obs} = \frac{k_\mathrm{int}}{P}.
        \label{eq:si-ll-ex2}
\end{align}
\end{subequations}
To avoid the assumption $k_\mathrm{cl} \gg k_\mathrm{op}$, one possibility is to assume pre-equilibrium between open and closed states whence motility ($k_\mathrm{op} + k_\mathrm{cl}$) is much faster than exchange ($k_\mathrm{int}$).
Then,
\begin{align}
    k_\mathrm{obs} = \frac{k_\mathrm{int}}{1+P}.
\label{eq:si-ll-pre-equilibrium}
\end{align}
Note that Eq.\! \ref{eq:si-ll-pre-equilibrium} holds for any $P$ and reduces to the EX2 case (Eq.\! \ref{eq:si-ll-ex2}) if $P\gg1$, i.e.\! if the condition $k_\mathrm{cl} \gg k_\mathrm{op}$ is reintroduced.
Conversely, if $P\ll1$, $k_\mathrm{obs} \simeq k_\mathrm{int}$.

\newpage

\section{Generalized Linderstr{\o}m-Lang model}

The generalized Linderstr{\o}m-Lang (GLL) model \begin{align}
        \ce{H_{cl}} \xrightleftharpoons[k_\mathrm{cl}]{k_\mathrm{op}} \ce{H_{op}} \xrightleftharpoons[k_\mathrm{back}]{k_\mathrm{forw}} \ce{D_{op}} \xrightleftharpoons[k_\mathrm{op}']{k_\mathrm{cl}'} \ce{D_{cl}},
\label{eq:si-gll}
\end{align}
is described by a system of ordinary differential equations $\dot{\mathbf{x}}(t) = \mathbf{K} \mathbf{x}(t)$, where $\mathbf{x}(t)$ is the state vector of the system, whose components are the (normalized) populations $H_\mathrm{cl}(t)$, $H_\mathrm{op}(t)$, $D_\mathrm{op}(t)$, $D_\mathrm{cl}(t)$, and
\begin{align}
    \mathbf{K} = \begin{bmatrix}
        - k_\mathrm{op} & + k_\mathrm{cl} & 0 & 0 \\
        + k_\mathrm{op} & - k_\mathrm{cl} - k_\mathrm{forw} & + k_\mathrm{back} & 0 \\
        0 & + k_\mathrm{forw} & - k_\mathrm{cl}' - k_\mathrm{back} & + k_\mathrm{op}' \\
        0 & 0 & + k_\mathrm{cl}' & - k_\mathrm{op}'
    \end{bmatrix}.
\label{eq:si-K-gll}
\end{align}
The natural state space for the vectors $\mathbf{x}(t)$ is \[ \Omega = \left\{ \mathbf{r} = \begin{bmatrix} r_1 & r_2 & \cdots & r_n \end{bmatrix}^\text{T} \text{ s.t. } r_k \in \mathbb{R}, r_k >0 \text{ for all } k=1,2,\dots,n, \text{ and } \sum_{i=1}^n r_k = 1 \right\}, \] also called the probability simplex. Here, $n=4$.

The stationary solution, i.e.\! satisfying $\mathbf{K} \mathbf{x}_\mathrm{eq} = \mathbf{0}$, is
\begin{align}
    \mathbf{x}_\mathrm{eq} = \frac{ \begin{bmatrix}
        K_\mathrm{back} P & K_\mathrm{back} & 1 & P'
    \end{bmatrix}^\text{T}}{K_\mathrm{back}(1+P) + (1+P')} = \frac{\begin{bmatrix}
        K_\mathrm{back} P & K_\mathrm{back} & 1 & P+\delta P
    \end{bmatrix}^\text{T}}{(1+K_\mathrm{back})(1+P) + \delta P},
\label{eq:si-xeq-delta}
\end{align}
where $P=k_\mathrm{cl}/k_\mathrm{op}$ and $P'=k_\mathrm{cl}'/k_\mathrm{op}' = P(1+\delta)$ are protection factors, and $K_\mathrm{back} = k_\mathrm{back}/k_\mathrm{forw}$ is the equilibrium constant of the back exchange reaction.
Existence of a stationary solution is always guaranteed for the model \eqref{eq:si-gll} because the equations of the system (or the rows of $\mathrm{K}$) are linearly dependent. This also guarantees that $\mathbf{K}$ has an eigenvalue $\lambda_0 = 0$, whose eigenvector is $\mathbf{v}_0 \propto \mathbf{x}_\mathrm{eq}$.
From Eq.\! \ref{eq:si-xeq-delta}, it results
\begin{align}
    \frac{D_\mathrm{eq}}{H_\mathrm{eq}} = \frac{1}{K_\mathrm{back}}\left( \frac{1+P'}{1+P} \right) = \frac{1}{K_\mathrm{back}}\left( 1 + \frac{\delta P}{1+P} \right) .
\label{eq:si-gll-DHeq}
\end{align}

\subsection{Exact solution}

The solution of the GLL model \eqref{eq:si-gll} can always be written in terms of the flow $\phi^t : \Omega \to \Omega$, which here is simply the matrix exponential.
Its action evolves an initial condition $\mathbf{x}(0)$ as $\mathbf{x}(t) = \phi^t \left( \mathbf{x}(0) \right) = \mathrm{e}^{\mathbf{K}t} \mathbf{x}(0)$.
Diagonalizing matrix $\mathbf{K}$, the solution can be written in terms of eigenvalues $\{\lambda\}$ and eigenvectors $\{\mathbf{v}_\lambda\}$ of $\mathbf{K}$ as \[ \mathbf{x}(t) = \sum_\lambda c_\lambda \mathrm{e}^{-\lambda t} \mathbf{v}_\lambda = \mathbf{x}_\mathrm{eq} + \sum_{\lambda \neq 0} c_\lambda \mathrm{e}^{-\lambda t} \mathbf{v}_\lambda, \] where $\{c_\lambda\}$ are coefficients that depend on initial condition $\mathbf{x}(0)$ and in the last equality the equilibrium state $\mathbf{x}_\mathrm{eq}$ has been evidenced.
$\mathbf{K}$ is a real tridiagonal matrix, i.e. it has the form
\begin{align}
    \begin{bmatrix}
    a_1 & b_1 & 0 & && 0\\
    c_1 & a_2 & b_2 \\
    0 & c_2 & \ddots & \ddots &  &\\ 
     & & \ddots & & \ddots\\
     & & & \ddots& \ddots & b_{n-1} \\
    0 & && & c_{n-1} & a_{n}
    \end{bmatrix}.
\label{eq:si-tridiagonal}
\end{align}
Because $\mathbf{K}$ (Eq. \ref{eq:si-K-gll}) is similar to a real symmetric tridiagonal matrix $\mathbf{S}$ with non-null diagonal and off-diagonal values, a theorem guarantees that all its eigenvalues are real and simple (i.e.\!, they are all distinct and associated to a unique eigenvector) \cite{parlett1998symmetric}.
The eigenvalues of $\mathbf{S}$ coincide with those of $\mathbf{K}$.
If $\mathbf{S} = \mathbf{D}^{-1} \mathbf{K} \mathbf{D}$, the eigenvector $\mathbf{v}_\lambda$ of $\mathbf{K}$ is given by $\mathbf{v}_\lambda = \mathbf{\mathbf{D} \mathbf{u}_\lambda}$, where $\mathbf{u}_\lambda$ is the eigenvector of $\mathbf{S}$ corresponding to the same eigenvalue $\lambda$.
$\mathbf{S}$ can be obtained for example by a diagonal matrix $\mathbf{D} = \mathrm{diag}(d_k)$ with $d_k = \sqrt{x_{\mathrm{eq},k}}$, where the $\{x_{\mathrm{eq},k}\}$ are the components of $\mathbf{x}_\mathrm{eq}$ (Eq.\! \ref{eq:si-xeq-delta}). The result is
\begin{align}
    \mathbf{S} =
    \begin{bmatrix}
    - k_\mathrm{op} & + \sqrt{k_\mathrm{op} k_\mathrm{cl}} & 0 & 0 \\
    + \sqrt{k_\mathrm{op} k_\mathrm{cl}} & - k_\mathrm{cl}-k_\mathrm{forw} & + \sqrt{k_\mathrm{forw} k_\mathrm{back}} & 0 \\
    0 & \sqrt{k_\mathrm{forw} k_\mathrm{back}} & - k_\mathrm{cl}' - k_\mathrm{back} & + \sqrt{k_\mathrm{op}' k_\mathrm{cl}'} \\
    0 & 0 & + \sqrt{k_\mathrm{op}' k_\mathrm{cl}'} & - k_\mathrm{op}'
    \end{bmatrix}.
\label{eq:si-S}
\end{align}
The desired eigenvalues $\lambda_k$ are the roots of the characteristic polynomial $p(\lambda) = \det (\lambda \mathbf{I} -  \mathbf{S})$.
For a $n\times n$ real tridiagonal matrix $\mathbf{T}$ with entries named as in Eq.\! \ref{eq:si-tridiagonal}, let $\mathbf{T}_m$ the $m \times m $ ($m\leq n$) principal submatrix of $\mathbf{T}$, obtained deleting rows and columns $m+1,m+2,\dots,n$ from $\mathbf{T}$.
The polynomials $p_m(\lambda) = \det(\lambda \mathbf{I} - \mathbf{T}_m)$ satisfy the recurrence relation
\begin{align}
    p_m(\lambda) = (\lambda - a_m) p_{m-1}(\lambda) - b_{m-1}c_{m-1} p_{m-2}(\lambda),
\label{eq:si-recurrence}
\end{align}
with initial conditions $p_0(\lambda) = 1$, $p_1(\lambda) = \lambda - a_1$.
Because by definition  $\mathbf{T} = \mathbf{T}_n$, the characteristic polynomial of $\mathbf{T}$ is $p_n(\lambda)$.
Dividing Eq.\! \ref{eq:si-recurrence} by $b_1 \cdots b_{n-1}$, a recursive relation for the eigenvectors is obtained:
\[ u_{\lambda,j } = w_\lambda \frac{p_{j-1} (\lambda)}{b_1 \cdots b_{j-1}}, \quad j=2,\dots,n,\]
where $u_{\lambda,j}$ is the $j$-th component of the eigenvector $\mathbf{u}_{\lambda}$ of $\mathbf{T}$, and $u_{\lambda,1} = w_\lambda$ is determined by the normalization as \[w_\lambda^{-1} = \sqrt{ \frac{p_{n-1}(\lambda)}{(b_1 \cdots b_{n-1})^2} \prod_{\lambda' \neq \lambda} (\lambda' - \lambda)}.\]

The characteristic polynomial of $\mathbf{S}$ (Eq.\! \ref{eq:si-S}) was computed recursively by Eq.\! \ref{eq:si-recurrence}:
\begin{align}
    p(\lambda) = \lambda \left( \lambda^3 + A \lambda^2 + B \lambda + C  \right),
\label{eq:si-cpol}
\end{align}
where
\begin{align*}
\begin{split}
        A =& k_\mathrm{op} + k_\mathrm{op}' + k_\mathrm{cl} + k_\mathrm{cl}' + k_\mathrm{forw} + k_\mathrm{back}, \\
        B = &k_\mathrm{op}k_\mathrm{op}' + k_\mathrm{op}k_\mathrm{cl}' + k_\mathrm{op}k_\mathrm{forw} + k_\mathrm{op}k_\mathrm{back}  +k_\mathrm{op}'k_\mathrm{cl} \\
        &+ k_\mathrm{op}'k_\mathrm{forw} + k_\mathrm{op}'k_\mathrm{back} + k_\mathrm{cl}k_\mathrm{cl}' + k_\mathrm{cl}k_\mathrm{back} + k_\mathrm{cl}'k_\mathrm{forw}, \\
        C =& k_\mathrm{op}k_\mathrm{op}'k_\mathrm{forw} + k_\mathrm{op}k_\mathrm{op}'k_\mathrm{back} + k_\mathrm{op}k_\mathrm{cl}'k_\mathrm{forw} + k_\mathrm{op}'k_\mathrm{cl}k_\mathrm{back}.
\end{split}
\label{eq:si-ABC}
\end{align*}
From Eq.\! \ref{eq:si-cpol}, it is evident that one eigenvalue is $\lambda=0$.
The other three are the roots of the polynomial $\lambda^3 + A\lambda^2 + B\lambda + C$.
These can be obtained transforming the equation in a depressed cubic $y^3 + py + q=0$, upon the change of variable $\lambda = y - A/3$, which is a particular case of Cardano's method. The coefficients are
\[ p=B-\frac{A^2}{3},\quad q=\frac{2A^3}{27}-\frac{AB}{3}+C, \]
and the solutions $y_k$ are given by $y_k = z_{+,k} + z_{-,k}$, where $z_{\pm,k}$ is the $k$-th root of
\[ \sqrt[3]{-\frac{q}{2} \pm \sqrt{ \Delta_\mathrm{C}}}, \quad \text{with} \, \Delta_\mathrm{C} = \left(\frac{q}{2}\right)^2 + \left(\frac{p}{3}\right)^2.\]
Because in this case eigenvalues are ensured to be real and distinct, $\Delta_\mathrm{C} < 0$ and one can use the trigonometric (cosine) form of Cardano to express the solutions:
\[ y_k = 2\sqrt{-\frac{p}{3}} \cos \left[ \frac{1}{3} \arccos\left( \frac{3q}{2p} \sqrt{-\frac{3}{p}} - \frac{2\pi(k-1)}{3} \right) \right],\quad k=1,2,3.\]
Eigenvalues can be then recovered as $\lambda_k = y_k - A/3$, and eigenvectors computed from the recursive method described above.
However, the expressions obtained (in terms of the rates) result intractable unless assumptions on the rates are introduced.


\subsection{Approximations}

\subsubsection{Steady state approximation}

The steady state approximation (SSA) can be applied to the GLL model \eqref{eq:si-gll} assuming that the open states \ce{H_{op}} and \ce{D_{op}} are short-lived intermediates that form a unique transition state \cite{kurzynski2006thermodynamic}.
This condition is satisfied if $k_\mathrm{cl} \gg k_\mathrm{op}$ and $k_\mathrm{cl}' \gg k_\mathrm{op}'$, i.e.\! if $P\gg1$ and $P'\gg 1$, thus represents a generalized version of the native approximation in the Linderstr{\o}m-Lang model.
The reaction \eqref{eq:si-gll} is simplified to an effective single reaction \cite{kurzynski2006thermodynamic}:
\begin{align}
    \ce{H} \xrightleftharpoons[k_-]{k_+} \ce{D},
\label{eq:si-gll-ssa}
\end{align}
where
\begin{subequations}
    \begin{align}
    k_+ &= \frac{k_\mathrm{op} k_\mathrm{forw}k_\mathrm{cl}'}{k_\mathrm{cl}k_\mathrm{back}+k_\mathrm{cl}k_\mathrm{cl}'+k_\mathrm{cl}'k_\mathrm{forw}}, \label{eq:si-gll-ssa-k+} \\
        k_- &= \frac{k_\mathrm{cl} k_\mathrm{back}k_\mathrm{op}'}{k_\mathrm{cl}k_\mathrm{back}+k_\mathrm{cl}k_\mathrm{cl}'+k_\mathrm{cl}'k_\mathrm{forw}}. \label{eq:si-gll-ssa-k-}
    \end{align}
\end{subequations}
The solution of the kinetics is
\begin{align}
    D(t) = \frac{1}{1+K} + \left( D_0 - \frac{1}{1+K}\right) \mathrm{e}^{-(k_++k_-)t},
\end{align}
where $K$ is the equilibrium constant of reaction \eqref{eq:si-gll-ssa},  $K=k_\mathrm{+}/k_-$, that coincides with the ratio $D_\mathrm{eq}/H_\mathrm{eq}$ of Eq.\! \ref{eq:si-gll-DHeq}.
One can rewrite Eqs.\! \ref{eq:si-gll-ssa-k+} and \ref{eq:si-gll-ssa-k-} as\cite{kurzynski2006thermodynamic}
\begin{subequations}
    \begin{align}
    k_+ &= \left[ \left( \frac{H_\mathrm{op,eq}}{H_\mathrm{cl,eq}} k_\mathrm{forw}\right)^{-1} + \left( k_\mathrm{op}\right)^{-1} + \left( \frac{D_\mathrm{eq}}{H_\mathrm{eq}} k_\mathrm{op}' \right)^{-1} \right]^{-1}, \label{eq:si-gll-ssa-k+inv} \\
    k_- &= \left[ \left( \frac{D_\mathrm{op,eq}}{D_\mathrm{cl,eq}} k_\mathrm{back}\right)^{-1} + \left( k_\mathrm{op}'\right)^{-1} + \left( \frac{H_\mathrm{eq}}{D_\mathrm{eq}} k_\mathrm{op} \right)^{-1} \right]^{-1}.
    \label{eq:si-gll-ssa-k-inv}
    \end{align}
\end{subequations}
Since by assumption $P \gg 1$, the equilibrium constant $K$ can be simplified to
\begin{align}
    K = \frac{1+\delta}{K_\mathrm{back}},
\label{eq:si-gll-ssa-K}
\end{align}
i.e.\! it is independent of $P$ and accounts for $P\neq P'$ by $\delta$.
Introducing the definitions of $P$, $P'$ and $K$, the rates $k_\pm$ become
\begin{subequations}
    \begin{align}
    (k_+)^{-1} &= \left( \frac{k_\mathrm{forw}}{P} \right)^{-1} + \left( k_\mathrm{op}\right)^{-1} + \left( \frac{1+\delta}{K_\mathrm{back}} \right)^{-1} (k_\mathrm{op}')^{-1}, \label{eq:si-gll-ssa-k+adj} \\
    (k_-)^{-1} &= \left( \frac{k_\mathrm{back}}{P'} \right)^{-1} + \left( k_\mathrm{op}'\right)^{-1} + \left( \frac{1+\delta}{K_\mathrm{back}} \right) (k_\mathrm{op})^{-1}.
    \label{eq:si-gll-ssa-k-adj}
    \end{align}
\end{subequations}
In Eqs. \ref{eq:si-gll-ssa-k+adj} and \ref{eq:si-gll-ssa-k-adj}, the first term indicates the time required to cross the transition state (which is assumed in equilibrium with the initial but not with the final state), the other two describe processes that restore equilibrium from the side of initial and final states.
If crossing the barrier is the limiting factor, the first components are the longest: \[ k_+ = \frac{k_\mathrm{forw}}{P}, \quad k_- = \frac{k_\mathrm{back}}{P'},
\]
which give an observed rate
\begin{align}
    k_\mathrm{obs} = \frac{k_\mathrm{int,mix}}{P} \left( 1  - \frac{K_\mathrm{back}}{1 + K_\mathrm{back}} \frac{\delta}{1 + \delta} \right),
\label{eq:si-gll-ssa-EX2}
\end{align}
where $k_\mathrm{int,mix} = k_\mathrm{forw} + k_\mathrm{back}$.
Conversely, in EX1, one has
\begin{align}
    k_\mathrm{obs} = k_\mathrm{op} \left( 1 - \frac{1-\xi}{K + \xi}\right),
\label{eq:si-gll-ssa-EX1}
\end{align}
where
\begin{align}
    \xi = \frac{k_\mathrm{op}}{k_\mathrm{op}'},
\label{eq:si-gll-ssa-xi}
\end{align} and $K$ is given by Eq.\! \ref{eq:si-gll-ssa-K}.
Eqs.\! \ref{eq:si-gll-ssa-EX2} and \ref{eq:si-gll-ssa-EX1} are formally analogous to the solutions of the Linderstr{\o}m-Lang model, Eqs.\! \ref{eq:si-ll-ex1} and \ref{eq:si-ll-ex2}, upon substitution of $k_\mathrm{int}$ by $k_\mathrm{int,mix}$ and introducing a factor that accounts for $k_\mathrm{cl}'\neq k_\mathrm{cl}$ and $k_\mathrm{op}'\neq k_\mathrm{op}$.
The LL case is evidently retrieved if $K_\mathrm{back} = 0$, i.e. $k_\mathrm{forw} = k_\mathrm{int}$ and $k_\mathrm{back} = 0$.

\subsubsection{Pre-equilibrium approximation}

A pre-equilibrium approximation applied to the GLL model \eqref{eq:si-gll} implies a separation of time scales:
if motility ($k_\mathrm{op}+k_\mathrm{cl}$ and $k_\mathrm{op}'+k_\mathrm{cl}'$) is much faster than exchange ($k_\mathrm{int,mix} = k_\mathrm{forw}+k_\mathrm{back}$),
\ce{H_{op}} and \ce{D_{op}} can be assumed to attain instantaneous equilibrium with their closed counterparts, thus, for any $t$, \[ \frac{H_\mathrm{op}(t)}{H_\mathrm{cl}(t)} = P, \; \frac{D_\mathrm{op}(t)}{D_\mathrm{cl}(t)} = P' \implies H_\mathrm{op}(t) = \frac{H(t)}{1+P},\; D_\mathrm{op}(t) = \frac{D(t)}{1+P'}.\]
Because of these relations, the study of the GLL model \eqref{eq:si-gll} reduces to a two-state reaction analogous to \eqref{eq:si-gll-ssa}.
Here, the resulting equation is
\begin{align}
\begin{split}
        \dot{D}(t) =& k_\mathrm{forw} H_\mathrm{op} (t) - k_\mathrm{back} D_\mathrm{op}(t) \\
        =& \frac{k_\mathrm{forw}}{1 + P} (1- D(t)) - \frac{k_\mathrm{back}}{1+P'} D(t) \\
        =& - \frac{k_\mathrm{int,mix}}{1 + P} \left( 1 - \frac{K_\mathrm{back}}{1+K_\mathrm{back}} \frac{\delta P}{1+P+\delta P} \right) D(t) + \frac{k_\mathrm{int,mix}}{1 + P} \frac{1}{1+K_\mathrm{back}},
\end{split}
\label{eq:si-gll-pre-diffeq}
\end{align}
hence
\begin{align}
    k_\mathrm{obs} = \frac{k_\mathrm{int,mix}}{1+P} \left( 1 - \frac{K_\mathrm{back}}{1+K_\mathrm{back}} \frac{\delta P}{1+ P + \delta P} \right).
\label{eq:si-gll-pre-kobs}
\end{align}
For $P \gg 1$, $\delta P/(1 + P + \delta P) \simeq \delta/(1+\delta)$, and Eq.\! \ref{eq:si-gll-pre-kobs} simplifies to the EX2 case (Eq.\! \ref{eq:si-gll-ssa-EX2}).

\subsubsection{No isotopic substitution effects}

For the generalized Linderstr{\o}m-Lang model that ignores the effects of isotopic substitution on stability, i.e.\! $k_\mathrm{cl}'=k_\mathrm{cl}$, $k_\mathrm{op}'=k_\mathrm{op}$ ($\delta = 0$),
\begin{align}
        \ce{H_{cl}} \xrightleftharpoons[k_\mathrm{cl}]{k_\mathrm{op}} \ce{H_{op}} \xrightleftharpoons[k_\mathrm{back}]{k_\mathrm{forw}} \ce{D_{op}} \xrightleftharpoons[k_\mathrm{op}]{k_\mathrm{cl}} \ce{D_{cl}},
\label{eq:si-gll-delta0}
\end{align}
eigenvalues $\{\lambda\}$ and (non-normalized) eigenvectors $\{\mathbf{v}_\lambda\}$ satisfying $\mathbf{K} \mathbf{v}_\lambda = \lambda \mathbf{v}_\lambda$ can be explicitly written as
\begin{subequations}\label{eq:eig-delta0}
\begin{align}
    &\begin{cases}
        \begin{split}
            \lambda_0 &= 0\\
            \mathbf{v}_0 &= \begin{bmatrix}
                PK_\mathrm{back} & K_\mathrm{back} & 1 & P
            \end{bmatrix}^\text{T}
        \end{split}\;,
    \end{cases}
    \label{eq:gll-delta0-lambda0}\\
    &\begin{cases}
        \begin{split}
            \lambda_1 &= -(k_\mathrm{op}+k_\mathrm{cl}) \\
            \mathbf{v}_1 &= \begin{bmatrix}
                K_\mathrm{back} & -K_\mathrm{back} & -1 & 1
            \end{bmatrix}^\text{T}
        \end{split}\;,
    \end{cases}
    \label{eq:gll-delta0-lambda1}\\
    &\begin{cases}
        \begin{split}
            \lambda_\pm &= -\frac{1}{2}\left( \gamma \pm \sqrt{\gamma^2 - 4 k_\mathrm{op} k_\mathrm{int,mix}} \right) \\
            \mathbf{v}_\pm &= \begin{bmatrix}
                -1 &
                -\frac{k_\mathrm{op}}{k_\mathrm{cl}}
                +\frac{\lambda_\pm}{k_\mathrm{cl}} &
                -\frac{k_\mathrm{op}}{k_\mathrm{cl}}
                -\frac{\lambda_\pm}{k_\mathrm{cl}} &
                1
            \end{bmatrix}^\text{T}
        \label{eq:gll-delta0-lambdapm}
        \end{split}\;,
    \end{cases}
\end{align}
\end{subequations}
where $ K_\mathrm{back} = k_\mathrm{back}/k_\mathrm{forw}$ is the equilibrium constant of the elementary back exchange reaction, $P = k_\mathrm{cl}/k_\mathrm{op}$ is the protection factor, and $ \gamma = k_\mathrm{op}+k_\mathrm{cl}+k_\mathrm{forw}+k_\mathrm{back}$.
The eigenvalue $\lambda_0$ is associated with the stationary solution
\begin{align}
    \mathbf{x}_\mathrm{eq} = \frac{\mathbf{v}_0}{(1+K_\mathrm{back})(1+P)},
\label{eq:si-xeq-delta0}
\end{align}
which predicts an equilibrium amide deuteration $D_\mathrm{eq}$ that depends only on $K_\mathrm{back}$, that is,
\begin{align}
    \frac{D_\mathrm{eq}}{H_\mathrm{eq}} = \frac{1}{K_\mathrm{back}}
\label{eq:si-gll-delta0-DHeq}.
\end{align}
Eigenvalue $\lambda_\mathrm{1}$ and eigenvector $\mathbf{v}_1$ describe no net exchange, rather encode motility, i.e.\! opening/closing dynamics.\cite{hvidt1966hydrogen}
Eigenvalues $\lambda_\pm$ and eigenvectors $\mathbf{v}_\pm$ describe exchange.
It is noted that results of Eqs. \ref{eq:gll-delta0-lambda0}, \ref{eq:si-xeq-delta0} and \ref{eq:si-gll-delta0-DHeq} are formally analogous to those that hold for the GLL model \eqref{eq:si-gll} for $\delta=0$.

The SSA applied to the model \eqref{eq:si-gll-delta0} gives again a two-state reaction as in \eqref{eq:si-gll-ssa}, with rates
\begin{subequations}
    \begin{align}
    k_+ &= \frac{k_\mathrm{op} k_\mathrm{forw}}{k_\mathrm{cl}+k_\mathrm{forw}+k_\mathrm{back}}, \label{eq:si-gll-ssa-k+0} \\
    k_- &= \frac{k_\mathrm{op} k_\mathrm{back}}{k_\mathrm{cl}+k_\mathrm{forw}+k_\mathrm{back}}. \label{eq:si-gll-ssa-k-0}
    \end{align}
\end{subequations}
In this case,
\begin{align}
    k_\mathrm{obs} = \frac{k_\mathrm{op} k_\mathrm{int,mix}}{k_\mathrm{cl}+k_\mathrm{int,mix}},
\label{eq:si-gll-0-ssa-kobs}
\end{align}
which is formally analogous to the native approximation of the Linderstr{\o}m-Lang model.
In the EX1 case, $k_\mathrm{cl} \ll k_\mathrm{int,mix}$, Eq.\! \ref{eq:si-gll-0-ssa-kobs} reduces to
\begin{align}
    k_\mathrm{obs} = k_\mathrm{op}.
\label{eq:si-gll-0-EX1}
\end{align}
For the EX2 case, $k_\mathrm{cl} \gg k_\mathrm{op}$, one finds
\begin{align}
    k_\mathrm{obs} = \frac{k_\mathrm{int,mix}}{P}
\label{eq:si-gll-0-EX2}
\end{align}
Note that Eqs.\! \ref{eq:si-gll-0-EX1} and \ref{eq:si-gll-0-EX2} coincide with Eqs.\! \ref{eq:si-gll-ssa-EX1} and \ref{eq:si-gll-ssa-EX2} for $\delta = 0$ and $\xi=1$, \textit{cfr} Eq.\! \ref{eq:si-gll-ssa-xi}, which are exactly the assumptions introduced the model \eqref{eq:si-gll-delta0}.

The pre-equilibrium approximation gives
\begin{align}
    k_\mathrm{obs} = \frac{k_\mathrm{int,mix}}{1+P},
\end{align}
consistently with Eq.\! \ref{eq:si-gll-pre-kobs} for the case $\delta = 0$.

\newpage

\section{HDX/NMR kinetics of exchange}

\begin{figure}[h!]
    \centering
    \setlength{\tabcolsep}{1pt}    
    \renewcommand{\arraystretch}{0} 

    {\includegraphics[width=0.25\textwidth]{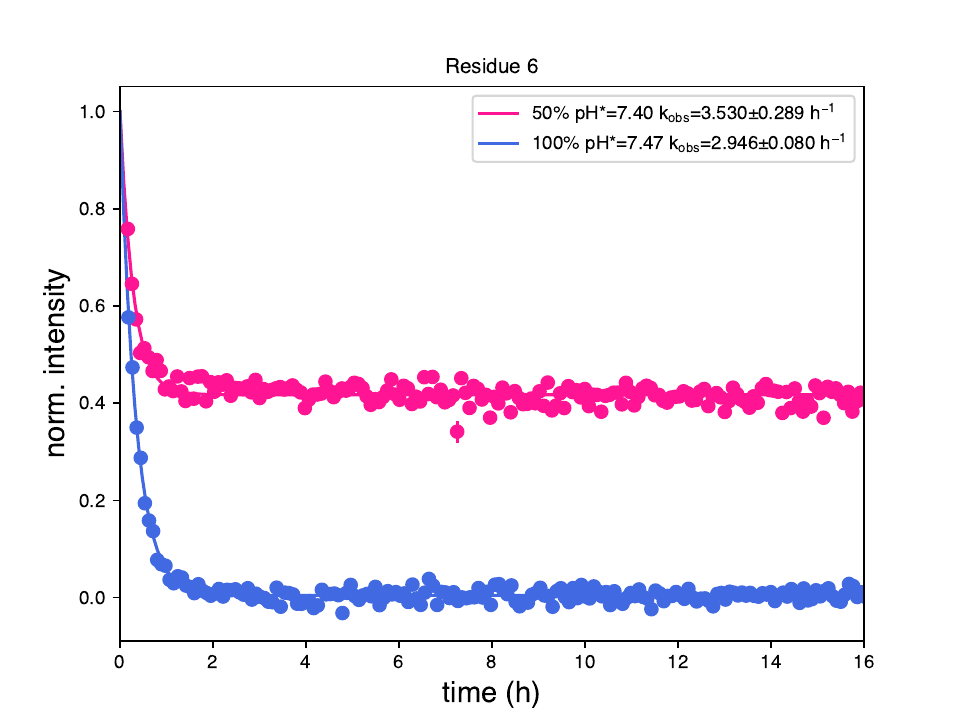}} \hfill
    {\includegraphics[width=0.25\textwidth]{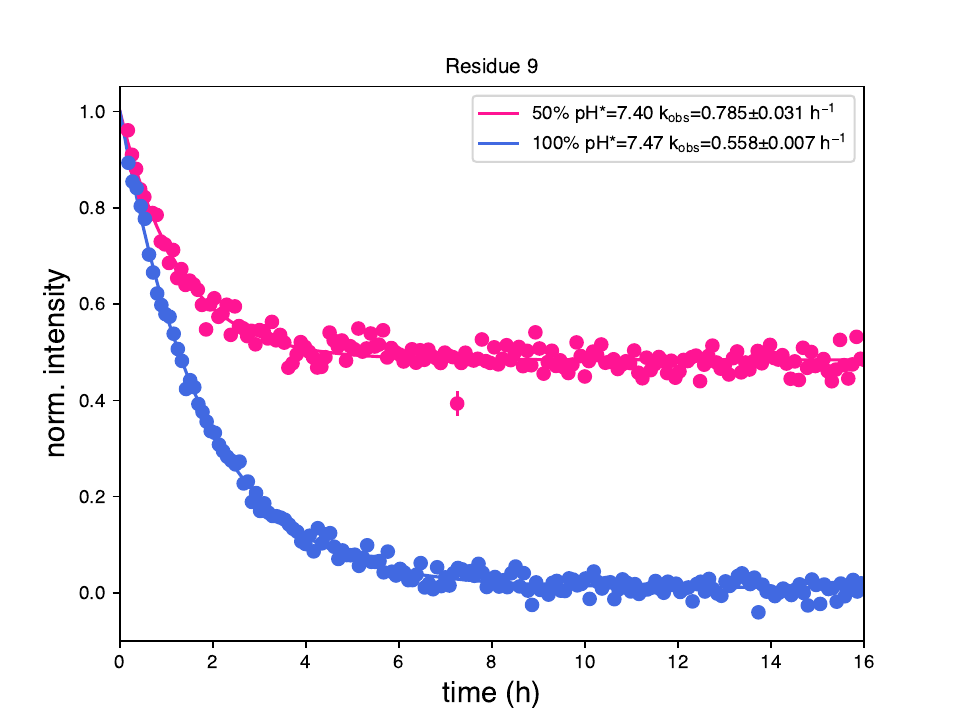}} \hfill
    {\includegraphics[width=0.25\textwidth]{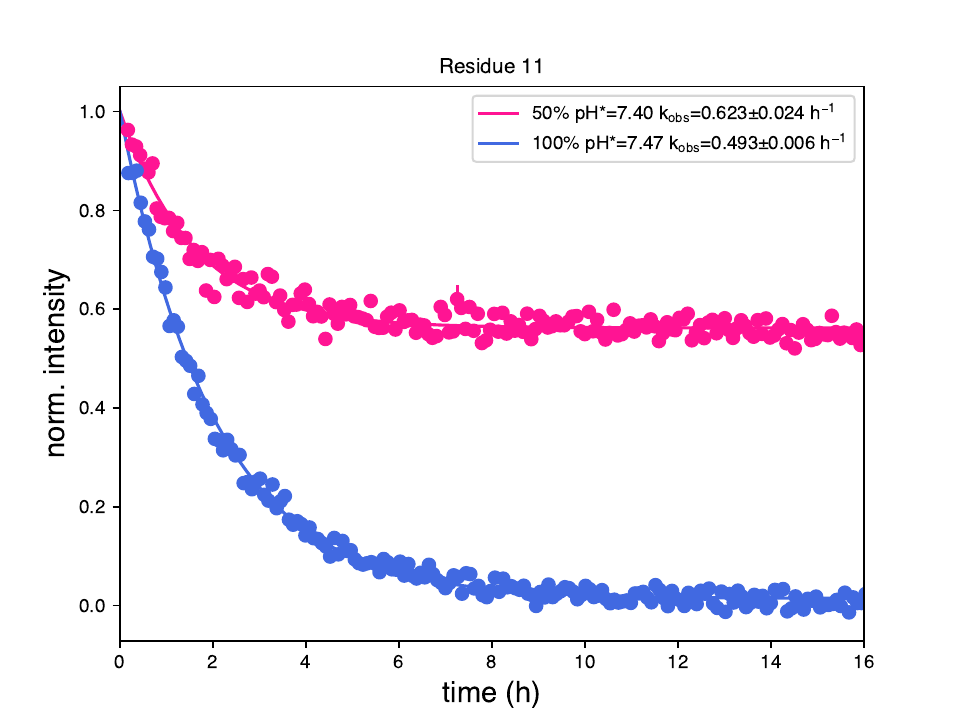}} \\
    {\includegraphics[width=0.25\textwidth]{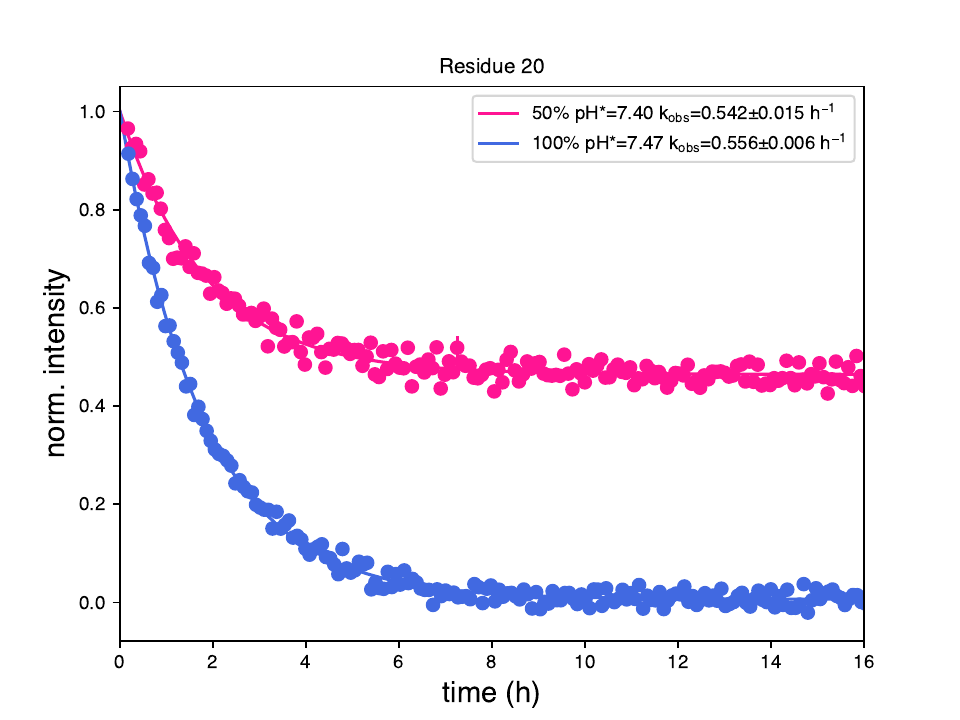}} \hfill
    {\includegraphics[width=0.25\textwidth]{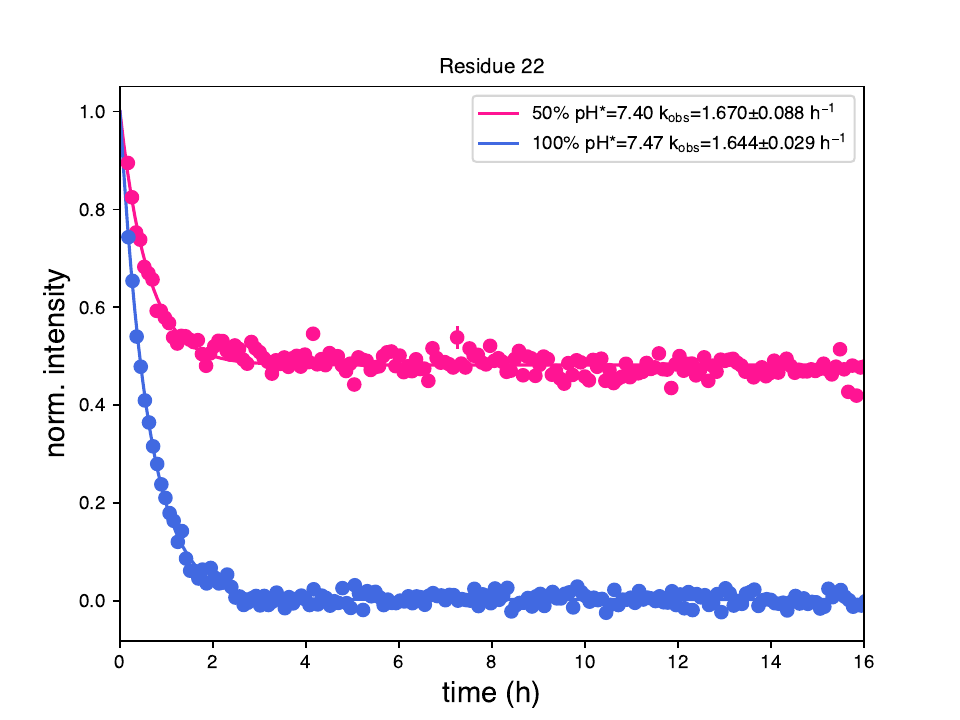}} \hfill
    {\includegraphics[width=0.25\textwidth]{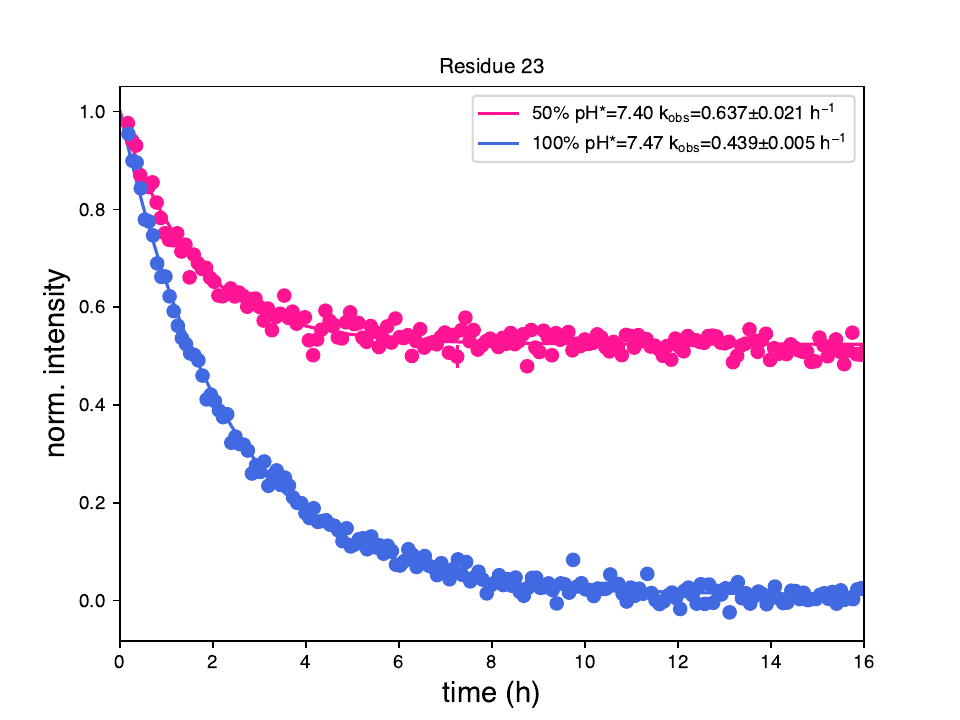}} \\
    {\includegraphics[width=0.25\textwidth]{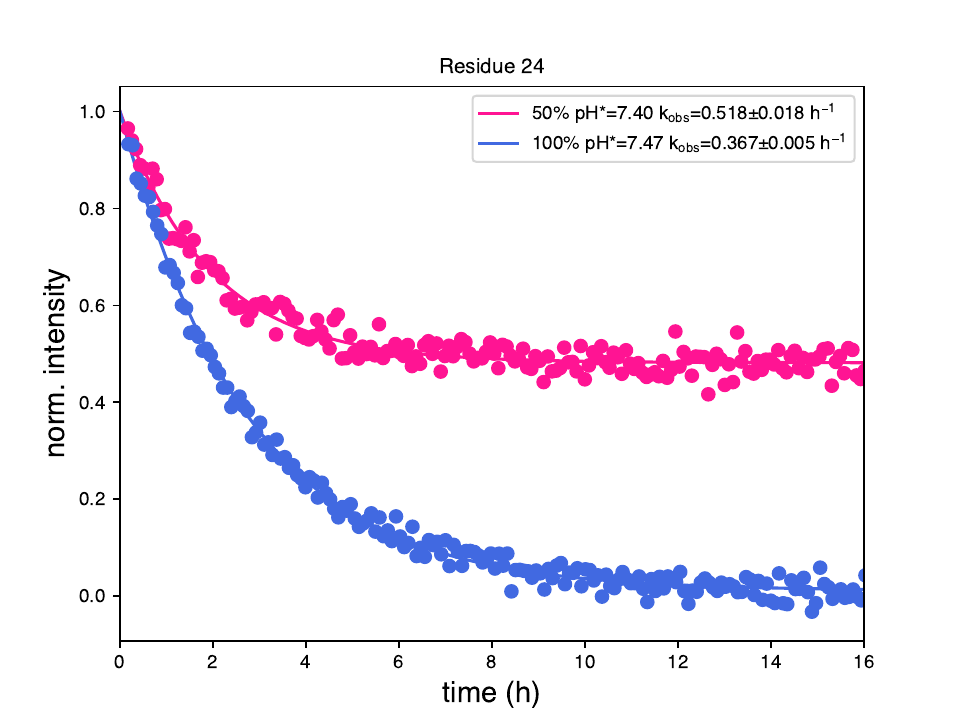}} \hfill
    {\includegraphics[width=0.25\textwidth]{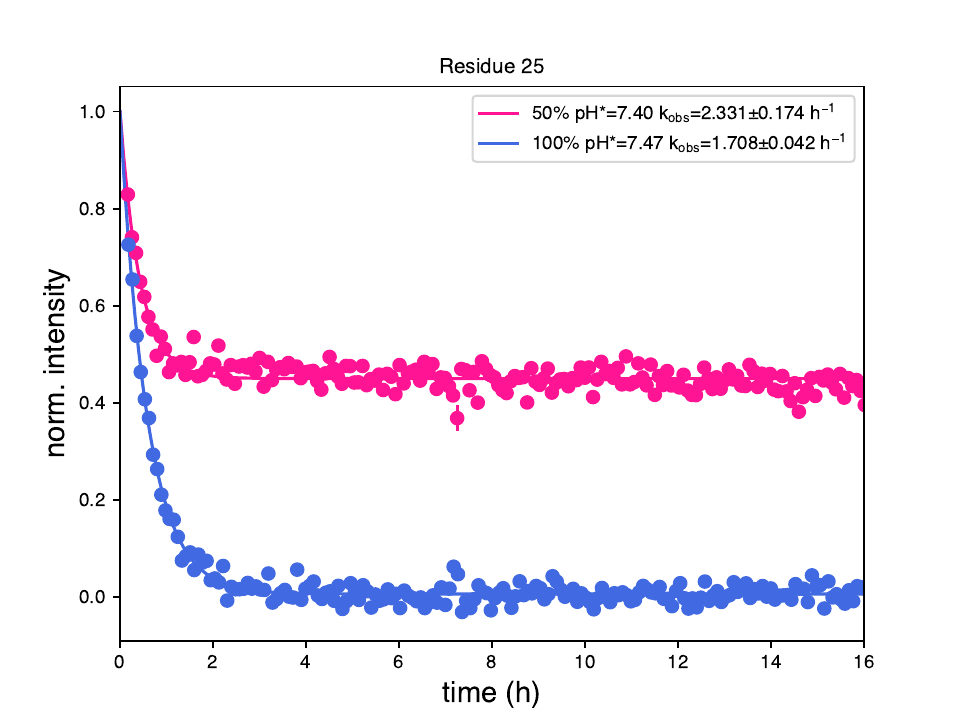}} \hfill
    {\includegraphics[width=0.25\textwidth]{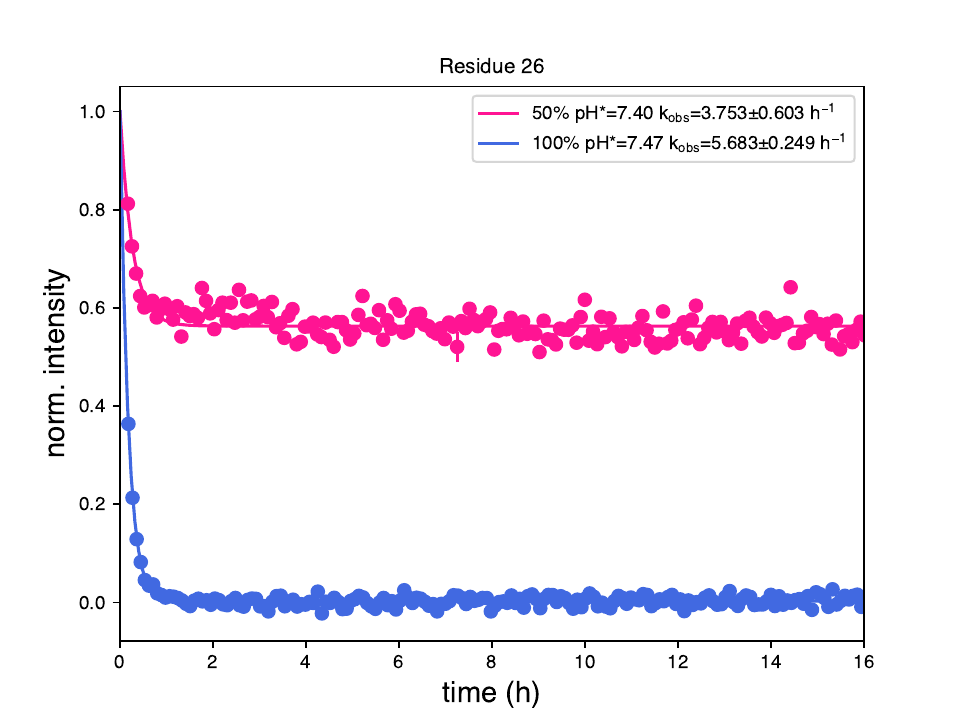}} \\
    {\includegraphics[width=0.25\textwidth]{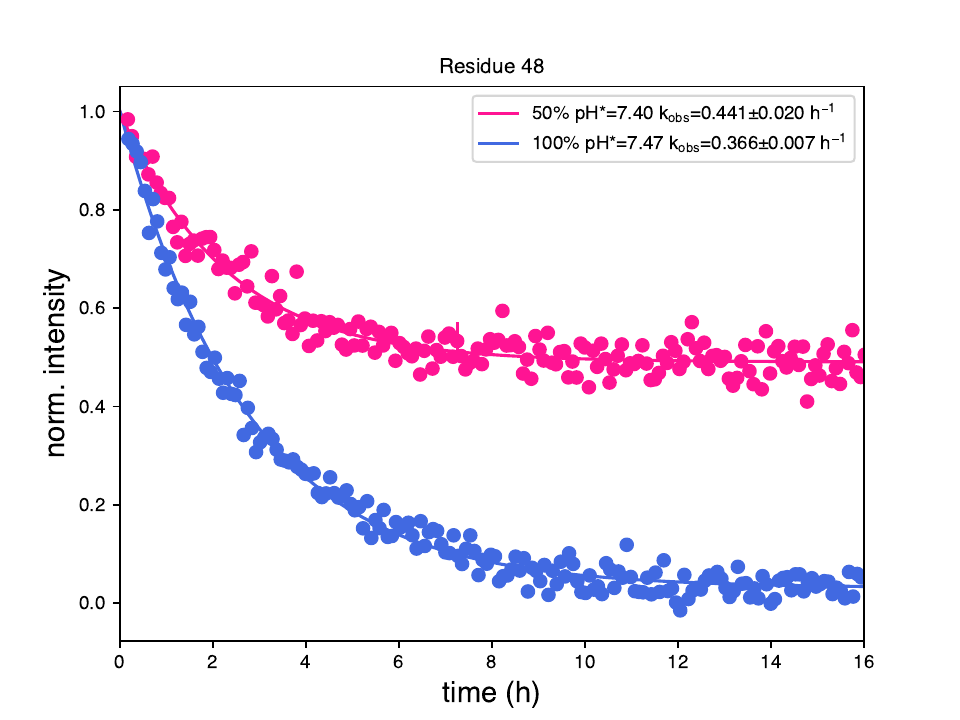}} \hfill
    {\includegraphics[width=0.25\textwidth]{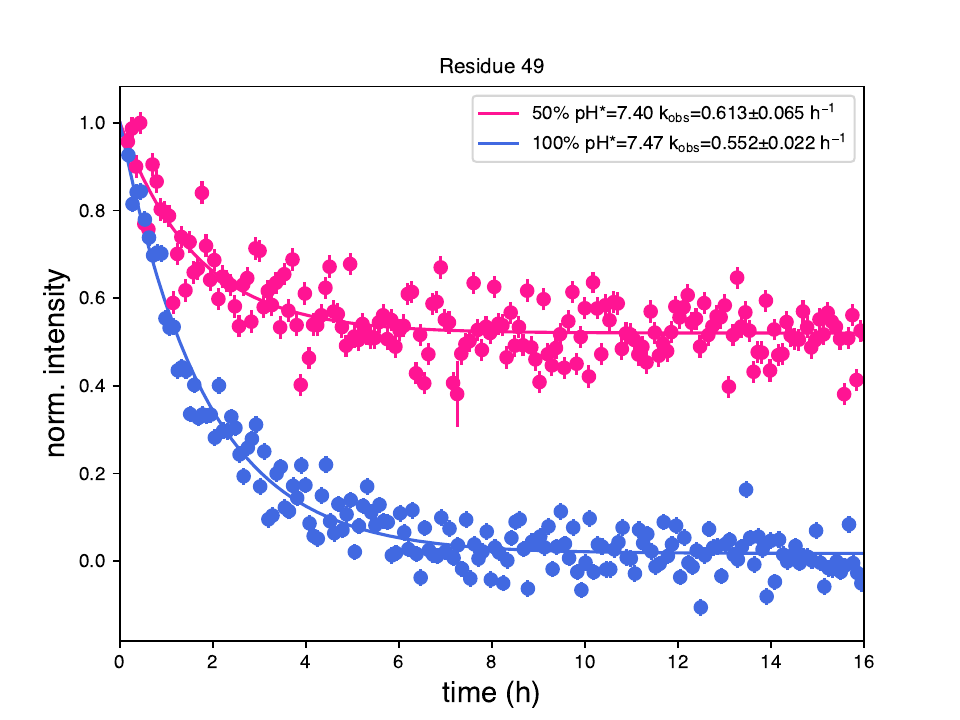}} \hfill
    {\includegraphics[width=0.25\textwidth]{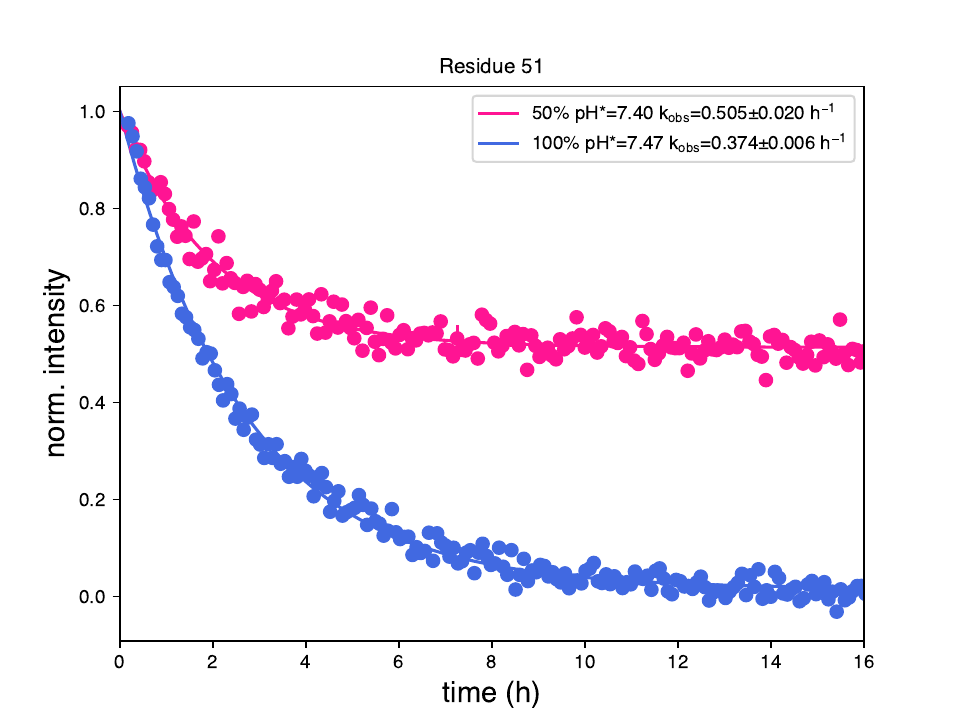}} \\
    {\includegraphics[width=0.25\textwidth]{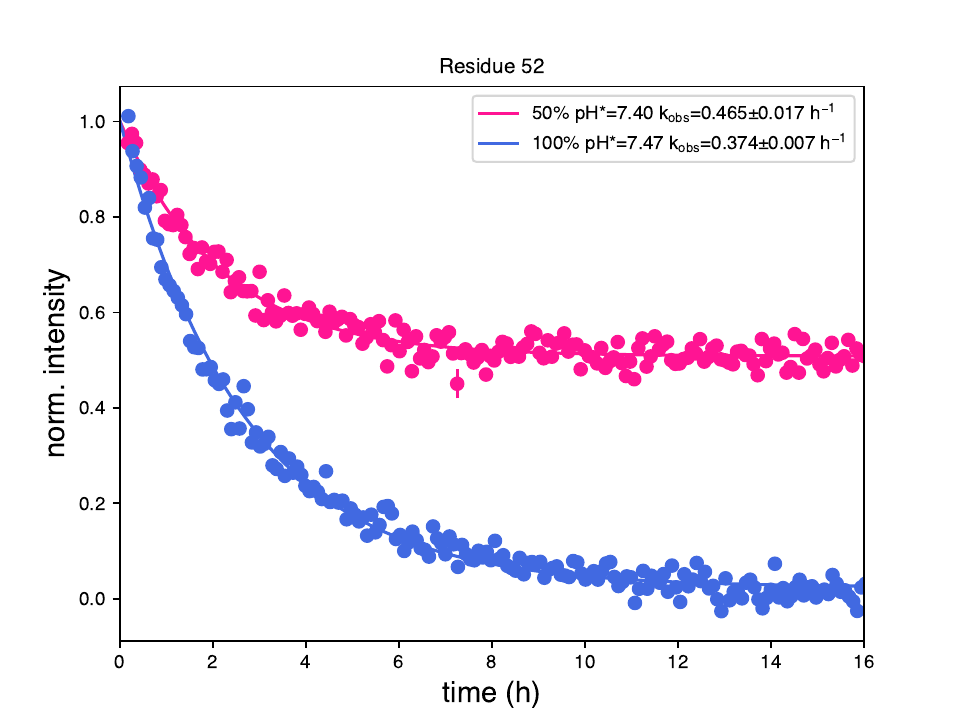}} \hfill
    {\includegraphics[width=0.25\textwidth]{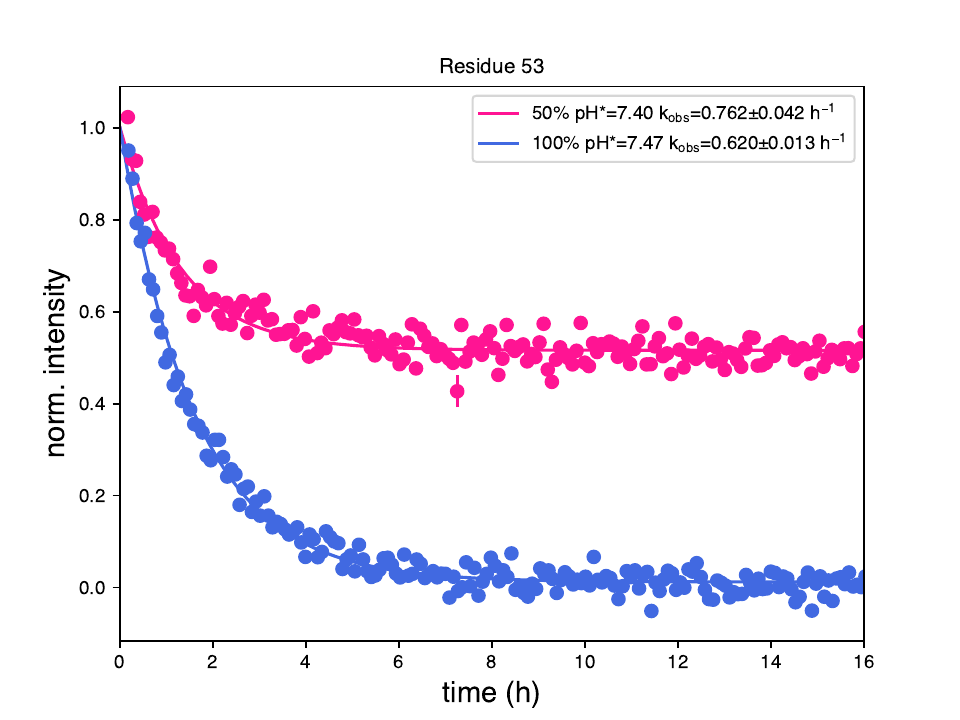}} \hfill
    {\includegraphics[width=0.25\textwidth]{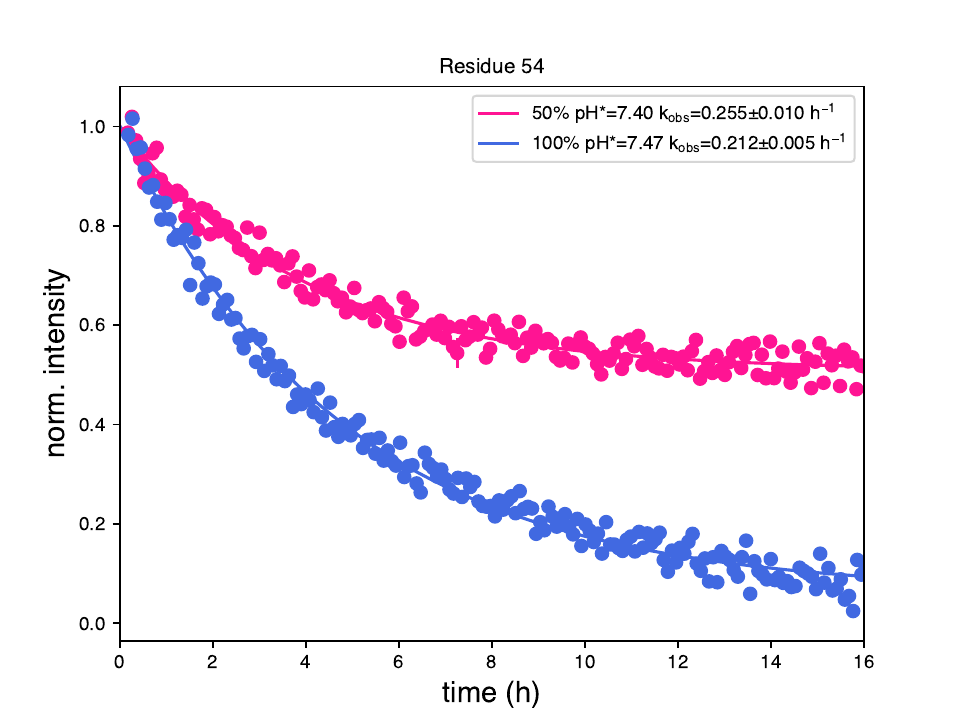}} \\
    {\includegraphics[width=0.25\textwidth]{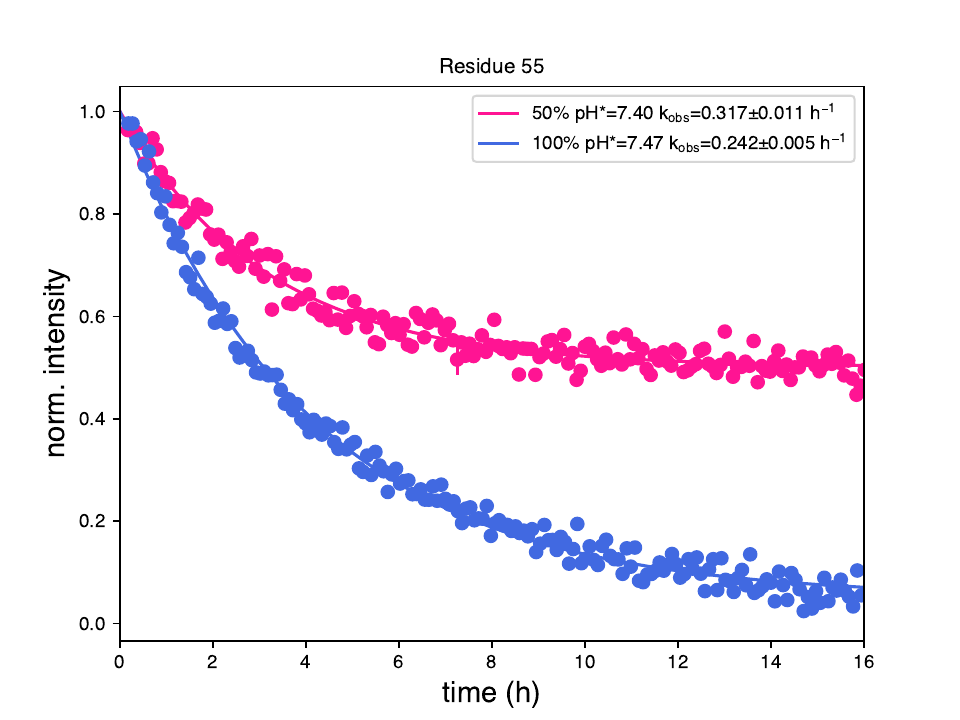}} \hfill
    {\includegraphics[width=0.25\textwidth]{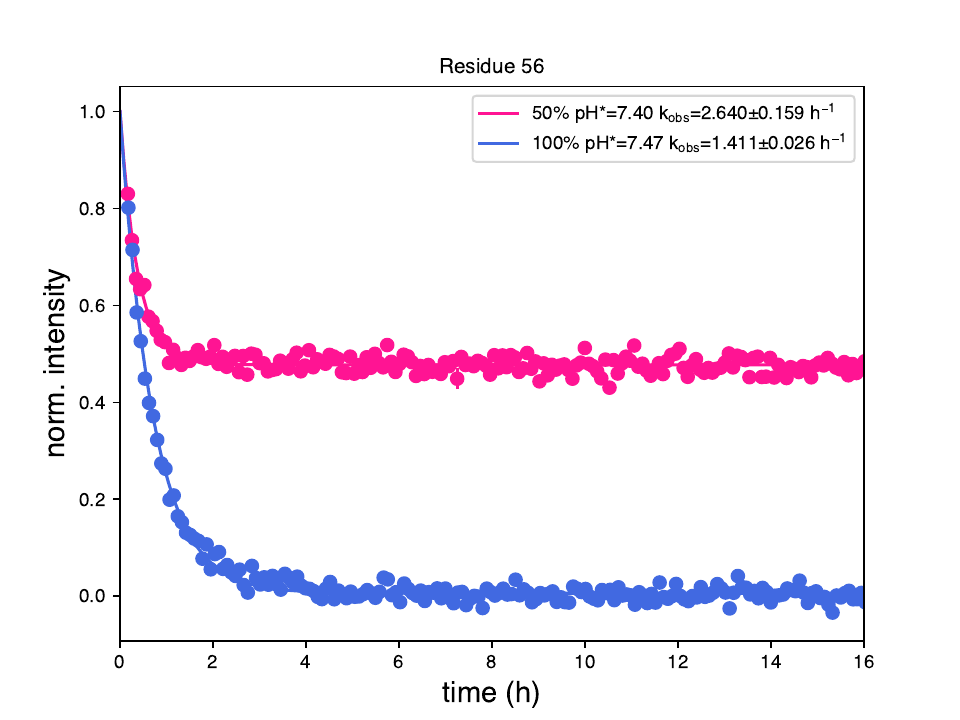}} \hfill
    {\includegraphics[width=0.25\textwidth]{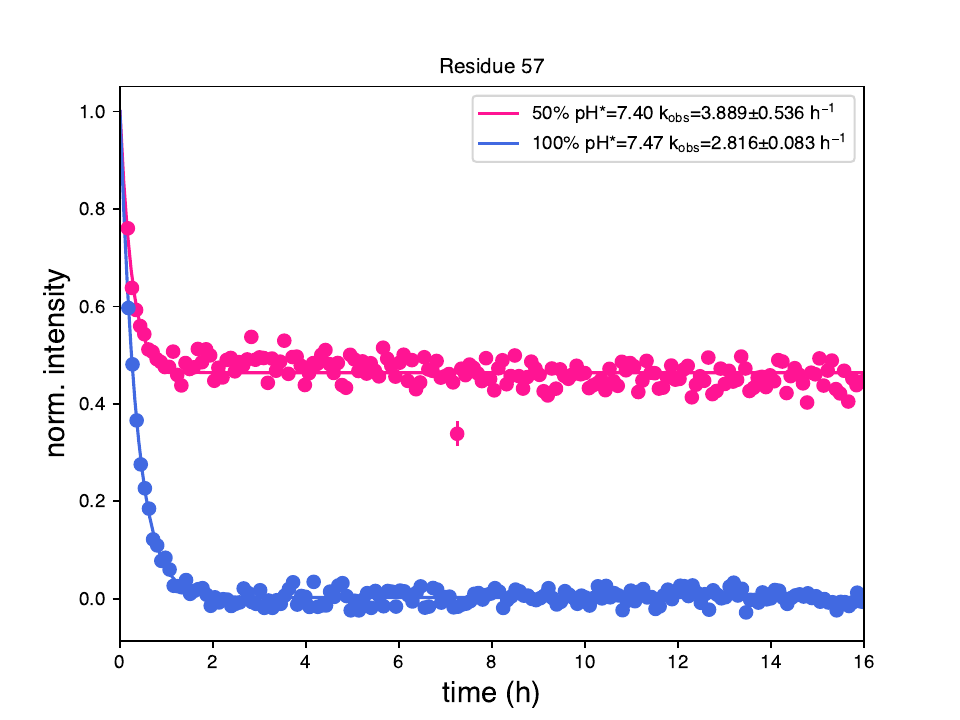}} \\

\caption{HDX-NMR experimental data (points) and exponential fit (line) for 18 amides of DNAJB1, in 50\% \ce{D2O} at $\mathrm{pH}^*=7.40$ (magenta) and 100\% \ce{D2O} at $\mathrm{pH}^*=7.47$ (blue).}
\end{figure}

\newpage

\bibliography{bib_si}